\documentclass[aps,prb, twocolumn, showpacs, amsmath, amssymb,superscriptaddress]{revtex4-1}

\usepackage{amssymb,amsfonts,amsmath} 
\usepackage{graphicx}
\usepackage{color}
\usepackage{longtable}

\usepackage{bbm}
\usepackage{ulem}
\usepackage{braket}

\usepackage{natbib}
\usepackage{url}
\usepackage[breaklinks=true]{hyperref}
\usepackage{mathtools}
\usepackage{subfigure}
\hypersetup{
        colorlinks=true,
        citecolor    = blue
}
\usepackage{bookmark}
\usepackage{epic}

\begin{document} 

\title{Exponential and power-law renormalization in phonon-assisted tunneling}

\author{A.\ Khedri}
\affiliation{Institut f{\"u}r Theorie der Statistischen Physik, RWTH Aachen University 
and JARA---Fundamentals of Future Information
Technology, 52056 Aachen, Germany}
\affiliation{Peter Gr\"unberg Institut and Institute for Advanced Simulation, Research Centre J\"ulich, 
52425 J\"ulich, Germany} 
\author{T.A.\ Costi}
\affiliation{Peter Gr\"unberg Institut and Institute for Advanced Simulation, Research Centre J\"ulich, 
52425 J\"ulich, Germany}
\author{V.\ Meden} 
\affiliation{Institut f{\"u}r Theorie der Statistischen Physik, RWTH Aachen University 
and JARA---Fundamentals of Future Information
Technology, 52056 Aachen, Germany}

\begin{abstract} 

We investigate the spinless Anderson-Holstein model routinely employed to describe the basic physics of 
phonon-assisted tunneling in molecular devices. Our focus is on small to intermediate electron-phonon 
coupling; we complement a recent strong coupling study 
[Phys.~Rev.~B {\bf 87}, 075319 (2013)]. The entire crossover from the antiadiabatic regime to the adiabatic 
one is considered. Our analysis using the essentially analytical functional renormalization group approach 
backed up by numerical renormalization group calculations goes beyond lowest order perturbation theory in the 
electron-phonon coupling. In particular, we provide an analytic expression for the effective tunneling 
coupling at particle-hole symmetry valid for all ratios of the bare tunnel coupling and the phonon 
frequency. It contains the exponential polaronic as well as the power-law renormalization in the 
electron-phonon interaction; the latter 
can be traced back to x-ray edgelike physics. In the antiadiabatic and the adiabatic limit this expression agrees with the known 
ones obtained by mapping to an effective interacting resonant level model and lowest order 
perturbation theory, respectively. Away from particle-hole symmetry, we discuss and compare
results from several approaches for the zero temperature electrical conductance of the model.
\end{abstract}

\date{\today} 
\maketitle

\section{Introduction}
\label{sec:intro}

Studying phonon effects on the spectral and transport properties of bulk electronic systems has 
a long history in condensed matter physics and is a topic of textbooks (see, e.g., Ref.~\onlinecite{Mahan90}). The development 
of molecular electronics led to a new twist to the electron-phonon problem. In such systems the molecular 
phonon modes only couple locally to a restricted number of relevant molecular electronic levels. The 
molecule is additionally coupled to electronic reservoirs via tunnel barriers with the tunnel coupling 
providing another energy scale. The basic physics of such systems can 
be obtained from model studies (see, e.g., Ref.~\onlinecite{Mitra04}). 

We here focus on one of the most elementary models of molecular electronics the so-called single-level 
spinless Anderson-Holstein model (SAHM) defined by the Hamiltonian
\begin{eqnarray}
H= H_{\rm lead} + H_{\rm mol} + H_{\rm coup} .
\label{eq:ham}
\end{eqnarray}  
The first term describes two (for simplicity) 
identical fermionic leads (reservoirs) with dispersion 
$\varepsilon_k$ 
\begin{eqnarray}
H_{\rm lead} = \sum_{\alpha=1}^{2} \sum_k \varepsilon_k c_{\alpha,k}^\dag c_{\alpha,k}.  
\label{eq:hamlead}
\end{eqnarray}
The second one is associated with the single-level molecule of energy $\epsilon_0$ coupled 
to a single phonon mode with frequency $\omega_0>0$ by the coupling constant $\lambda \geq 0$
\begin{eqnarray}
H_{\rm mol} = \epsilon_0 d^\dag d + \omega_0 b^\dag b + \lambda d^\dag d (b^\dag +b)  
\label{eq:hammol}
\end{eqnarray}
and the third models the molecule-reservoir tunnel coupling of amplitude $t$ 
\begin{eqnarray}
H_{\rm coup} = \frac{t}{\sqrt{N_{\rm sites}}} \sum_{\alpha=1}^{2} \sum_k \left( d^\dag c_{\alpha,k} + \mbox{H.c.}  \right) .
\label{eq:hamcoup}
\end{eqnarray}
Here $N_{\rm sites} \in {\mathbb N}$ denotes the number of lattice sites in each of the leads. 
Considering low temperatures this Hamiltonian shows intriguing many-body physics. 
We will discuss that this even holds in the limit of small to intermediate electron-phonon 
coupling $\lambda$ (for the reference scale, see below) on which we 
focus. This allows us to gain analytical results. 

Aiming at different goals which range from fundamental insights into correlation physics 
(e.g. the Kondo effect) to the explanation of experimental data the equilibrium physics of 
this model and its spinful variant was studied using a variety of approximate analytical as 
well as numerical methods.\cite{Mitra04,Sherrington75,Hewson79,Hewson80,Hewson02,Cornaglia04,Paaske05,Galperin07} 
Recently the attention shifted towards the nonequilibrium properties either 
in a bias-voltage driven steady state\cite{Hyldgaard94,Koenig96,Braig03,Koch05,Han06,*Han10,Muehlbacher08,Schiro09,Koch11,Huetzen12,Jovchev13,RouraBas13,Laakso14} 
or even considering the relaxation dynamics.\cite{Albrecht13} The former type of nonequilibrium studies led 
to a better understanding of the Franck-Condon blockade which was also observed in a molecular 
electronics experiment.\cite{Leturcq09} However, we here consider the equilibrium
properties (including linear transport), focus on the low-temperature correlation physics at small 
to intermediate $\lambda$, and this way fill a gap in our understanding of the model. 

Correlation effects are most prominent if the level energy $\epsilon_0$ is taken to be 
$\lambda^2/\omega_0 =E_{\rm p}$, the polaronic shift,\cite{Mahan90} for which the Hamiltonian 
Eqs.~(\ref{eq:ham})-(\ref{eq:hamcoup}) becomes particle-hole symmetric. Later on we will also consider 
general $\epsilon_0$ but for the following discussion consider this particle-hole symmetric 
point. 

In the  antiadiabatic limit $\Gamma \ll \omega_0$, with the (bare) tunnel coupling 
$\Gamma= 2 \pi \rho_{\rm lead} t^2$, as well as the complementary adiabatic regime 
$\Gamma \gg \omega_0$ the physics is rather well understood.
Here $\rho_{\rm lead}$ denotes the (assumed to be) constant lead density of states 
(wide-band limit; see below). Deep in the adiabatic regime $\Gamma \gg \omega_0$ the phonon is 
too slow to respond to the fermionic tunneling events which occur with high frequency; 
the effect of the phonon on the properties of the fermionic (sub-)\-system is minor. 
Consistently perturbation theory in $\lambda$ can even be used for (fairly) large electron-phonon 
couplings as it turns out that the expansion parameter is $E_{\rm p}/\Gamma$.  
In the antiadiabatic regime the phonon time scale $1/\omega_0$ is much smaller than the 
fermionic dwell time $1/\Gamma$ and the phonon can efficiently respond to hopping events. 
A polaron forms which leads to the well known suppression of the tunneling rate 
$\sim \exp{\left\{-(\lambda/\omega_0)^2\right\}}$ (polaronic suppression).\cite{Lang62,Mahan90} 

In the inspiring recent work by Eidelstein, Goberman, and 
Schiller\cite{Eidelstein13} this picture was refined in the antiadiabatic limit and 
complemented by results obtained in the crossover regime from antiadiabatic to adiabatic. 
Employing a Schrieffer-Wolff-like mapping of the SAHM to the interacting 
resonant level model (IRLM) and borrowing  established results for this the authors showed 
that the exponential suppression in the antiadiabatic regime is merely the zeroth order 
term in an expansion in $\Gamma/\omega_0$. In the limits of weak $\lambda \ll \omega_0$ 
and strong  $\lambda \gg \omega_0$ electron-phonon coupling one finds for the 
renormalized effective tunneling rate
\begin{eqnarray}
\frac{\Gamma_{\rm eff}^{\rm IRLM}}{\omega_0} = \left[ \frac{\Gamma}{\omega_0} e^{-(\lambda/\omega_0)^2} \right]^{1-\frac{4}{\pi} 
\frac{\Gamma}{\omega_0} u_\lambda} ,
\label{eq:Gammaeff}
\end{eqnarray}
with
\begin{eqnarray}
u_{\lambda} = \left\{ 
\begin{array}{cc}
\left( \frac{\lambda}{\omega_0} \right)^2 & \mbox{for} \; \lambda \ll \omega_0 \\
\left( \frac{\lambda}{\omega_0} \right)^{-2} & \mbox{for} \;  \lambda \gg \omega_0 
\end{array}
\right. .
\label{eq:ulambdadef}
\end{eqnarray}
An analytic expression for the function $u_\lambda$ beyond these two limits can be 
found in Ref.~\onlinecite{Eidelstein13}.    
Within the IRLM the power-law renormalization with argument $\Gamma/\omega_0$ can be traced back 
to x-ray edgelike physics.\cite{Schlottmann80,*Schlottmann82} The crossover from antiadiabatic to adiabatic behavior was 
studied using the numerical renormalization group (NRG) focusing on {\it large} electron-phonon 
couplings $\lambda \gg \omega_0$. It was shown that an extended antiadiabatic regime exists 
in which $\Gamma \ll \omega_0$ does no longer hold but the low-energy physics of the SAHM 
is still described by an effective IRLM. The crossover to the adiabatic regime sets  
in only for $\Gamma_{\rm eff} \approx \omega_0$ and the latter is eventually reached for 
$\Gamma \gtrapprox E_{\rm p}$.  

Our study is complementary to that of Ref. \onlinecite{Eidelstein13} as we consider the entire crossover in the 
limit of {\it small to intermediate} $\lambda$. Using an approximate functional renormalization group (FRG) 
approach which is controlled for such electron-phonon couplings we provide an {\it analytic expression}
of $\Gamma_{\rm eff}$ for {\it all} $\Gamma/\omega_0$. It contains the combined exponential and power-law 
renormalization of Eq.~(\ref{eq:Gammaeff}) in the antiadiabatic limit as well as the perturbative one 
obtained for $E_{\rm p} \ll \Gamma$. A comparison with NRG data shows that it provides a 
good approximation for all $\Gamma/\omega_0$. We discuss the limits of the mapping to the IRLM in 
the antiadiabatic regime. In addition, we discuss transport and spectral properties away from 
particle-hole symmetry $\epsilon_0 \neq E_{\rm p}$. In a followup
paper,\cite{Khedri17b} we extend the present work to investigate, within the NRG approach, the finite temperature linear  thermoelectric properties of the SAHM, comparing our results, where possible, with corresponding FRG calculations at finite temperature. 
       
The remainder of the paper is structured as follows. In Sect.~\ref{sec:method} we introduce our 
nonperturbative FRG approach to the SAHM. We consider the lowest order truncation\cite{Metzner12} 
which is controlled for small to intermediate electron-phonon coupling; being a single-particle term 
fermionic tunneling is considered to all orders. The coupled flow equations for the complex-valued 
self-energy are derived. Due to retardation effects the self-energy 
is frequency dependent. We discuss the relation between lowest order truncated FRG and first order 
perturbation theory in $\lambda^2$. Our NRG approach is briefly summarized. The results section 
\ref{sec:results} contains several subsections. In the first we derive a simplified flow equation 
for the imaginary part of the self-energy at particle-hole symmetry; the real part vanishes. 
This equation can be solved analytically for all $\Gamma/\omega_0$; from the self-energy 
$\Gamma_{\rm eff}$ can be computed. 
In the second subsection the renormalized tunneling rate obtained from the simplified equation is compared 
to the one derived from the numerical solution of the full lowest order flow equation as well as to 
$\Gamma_{\rm eff}$ determined from NRG. For $\lambda/\omega_0 \lessapprox 1$ the agreement is very good 
for all $\Gamma/\omega_0$. Finally, we consider the effects of
particle-hole asymmetry on the $T=0$
linear electrical conductance, comparing perturbation theory, FRG
and NRG results.
We conclude with a brief summary and outlook 
in Sec.~\ref{sec:summary}. Details of the implementation of the numerical solution of 
the full lowest order FRG flow equations, the convergence of the NRG
results with the number of phonon states, and a comparison of NRG
spectral functions with lowest order perturbation theory results are given in the Appendices.

\section{Methods}
\label{sec:method}
              
\subsection{General considerations}
\label{subsec:general}

Before introducing our methods we further characterize the model and summarize general properties. 

We assume that the fermionic leads feature particle-hole symmetric bands, that is that wave 
numbers come in pairs such that $\varepsilon_{k'} = - \varepsilon_{k}$.  
Under the transformation $d^\dag \to d$, $c_{\alpha,k}^\dag \to -c_{\alpha,k'}$, and
$b \to -b-\lambda/\omega_0$  the Hamiltonian Eqs.~(\ref{eq:ham})-(\ref{eq:hamcoup}) then 
becomes invariant provided the molecular dot level energy is chosen as 
$\epsilon_0=E_{\rm p}\equiv \lambda^2/\omega_0$.
One obtains $n_{\rm d}\left( \epsilon_0 \right) = \left< d^\dag d \right> = 1- n_{\rm d}\left( E_{\rm p} - 
\left[ \epsilon_0 - E_{\rm p}\right] \right)$ such that $\epsilon_0 = E_{\rm p}$ corresponds 
to half filling of the molecular level.
This defines the particle-hole symmetric point of the model. 
The quantity $\epsilon_0-E_p$, which controls the charge on the molecular dot, can be taken as the gate voltage on the dot.

As we are not interested in effects of details of the fermionic bands we take 
the so-called wide band limit and consider structureless reservoirs with constant density of states $\rho_{\rm lead}(\omega) = \rho_{\rm lead}$ for 
$\omega \in [-D,D]$, with the band width $2D$; it vanishes outside this energy interval. 
Integrating out the leads produces a reservoir contribution to the molecular 
self-energy  of the form $\Sigma_{\rm res} (i \xi_n) = -i \Gamma \mbox{sgn}\, (\xi_n)$, 
$\Gamma = 2 \pi t^2 \rho_{\rm lead}$. Here $\xi_n$ denotes a fermionic Matsubara frequency. 
For $\lambda=0$ the  single-particle Green function of the molecular level is then given by 
\begin{eqnarray}
G_{\rm mol}^0 (i \xi_n) = \left[ i \xi_n - \epsilon_0 + i \Gamma  \mbox{sgn}\, (\xi_n) \right]^{-1}
\label{eq:G0}
\end{eqnarray}    
and the corresponding spectral function is a Lorentzian of width $\Gamma$.   

\subsection{The functional RG}
\label{subsec:frg}

To set up our FRG approach following the standard procedure\cite{Metzner12} we integrate out the 
phonons in a functional integral approach to the many-body problem (see, e.g., Ref.~\onlinecite{Laakso14}). 
This way we end up 
with a purely fermionic action with a local ``on-molecule'', attractive, and retarded (frequency dependent) 
two-particle interaction of the form
\begin{eqnarray}
U(i \nu_n) = - \frac{2 \omega_0 \lambda^2}{\nu_n^2 + \omega_0^2} .
\label{eq:interact}
\end{eqnarray}
Here $\nu_n$ denotes a bosonic Matsubara frequency.
For this action we employ the FRG in its lowest order truncation with a bare two-particle 
vertex and a flowing self-energy.\cite{Metzner12} It is controlled for small 
to intermediate $\lambda$ but due to resummation of certain classes of diagrams inherent to the
RG procedure goes beyond simple perturbation theory. In particular, it was shown that 
this truncation captures the power-law renormalization of the tunnel coupling in the IRLM 
with an exponent which agrees with the exact one to leading order in the two-particle 
interaction.\cite{Karrasch10} As already mentioned in the introduction this piece of 
renormalization physics will also become essential in the antiadiabatic limit of the SAHM.
Further justification of our approximation will be given a posteori by comparing to the 
exact result in the antiadiabatic limit as well as to NRG results for general $\Gamma/\omega_0$.  

In contrast to earlier applications of 
lowest order FRG to correlated quantum dots\cite{Karrasch06,Karrasch10} the self-energy 
acquires a frequency dependence via the frequency dependence of the fermionic interaction. 
The present study must also be contrasted to an earlier work in which the
Anderson-Holstein model {\it with spin} was studied employing FRG.\cite{Laakso14,Kennes14} In 
this the focus was on the Kondo physics in the presence of a local phonon mode (mainly in 
bias-voltage driven nonequilibrium) which requires a truncation of the FRG equations to 
{\it higher} order.  

From now on we consider the zero temperature limit in which the Matsubara frequency 
becomes continuous. In our scheme the RG cutoff $\Lambda$ is introduced via this 
frequency. In the functional integral representation of the quantum 
many-body problem we replace 
the reservoir-dressed noninteracting molecular propagator Eq.~(\ref{eq:G0}) by 
$G_ {\rm mol}^{0,\Lambda}(i \nu) = G^0_{\rm mol}(i \nu) \Theta(|\nu| - \Lambda)$. Initially we take 
$\Lambda \to \infty$ to suppress any free propagation. The action is thus purely given by the 
interaction Eq.~(\ref{eq:interact}). The propagation is now turned on successively by sending 
$\Lambda$ to $0$; at $\Lambda=0$ the cutoff-free problem is recovered. This procedure
avoids logarithmic divergencies which might appear in a single-step perturbative  
treatment (see, e.g., Ref.~\onlinecite{Karrasch10} for the IRLM). Employing the generating functional
of the one-particle irreducible vertex functions and replacing the flowing effective two-particle 
interaction by the bare one  this procedure boils down to a set of coupled differential flow equations
for the self-energy.\cite{Metzner12} 
In the present case they read 
\begin{align}
& \partial_{\Lambda} \epsilon^\Lambda (i \nu) =  - \frac{2 E_{\rm p}}{\pi} 
\frac{ \epsilon^\Lambda (i \Lambda)}{\left[\Lambda + \Gamma - \gamma^{\Lambda}(i \Lambda) \right]^2 + 
\left[ \epsilon^\Lambda(i \Lambda) \right]^2} \nonumber \\
& + \frac{1}{\pi} \frac{ \omega_0 \lambda^2}{(\nu - \Lambda)^2+\omega_0^2} 
\frac{ \epsilon^\Lambda (i \Lambda)}{\left[\Lambda + \Gamma - \gamma^{\Lambda}(i \Lambda) \right]^2 + 
\left[ \epsilon^\Lambda(i \Lambda) \right]^2}  \nonumber \\
& + \frac{1}{\pi} \frac{ \omega_0 \lambda^2}{(\nu + \Lambda)^2+\omega_0^2} 
\frac{ \epsilon^\Lambda (-i \Lambda)}{\left[-\Lambda - \Gamma - \gamma^{\Lambda}(-i \Lambda) \right]^2 + 
\left[ \epsilon^\Lambda(-i \Lambda) \right]^2} , 
\label{eq:flowepsilon} 
\\
& \partial_{\Lambda}  \gamma^\Lambda (i \nu) =  \nonumber \\
& -  \frac{1}{\pi} 
\frac{ \omega_0 \lambda^2}{(\nu - \Lambda)^2+\omega_0^2}
\frac{ - \Lambda - \Gamma + \gamma^\Lambda(i \Lambda)}{\left[\Lambda + 
\Gamma - \gamma^{\Lambda}(i \Lambda) \right]^2 + 
\left[ \epsilon^\Lambda(i \Lambda) \right]^2}  \nonumber \\
&- \frac{1}{\pi} 
\frac{ \omega_0 \lambda^2}{(\nu + \Lambda)^2+\omega_0^2}
\frac{  \Lambda + \Gamma + \gamma^\Lambda(-i \Lambda)}{\left[-\Lambda - 
\Gamma - \gamma^{\Lambda}(-i \Lambda) \right]^2 + 
\left[ \epsilon^\Lambda(-i \Lambda) \right]^2} ,
\label{eq:flowgamma}
\end{align}
with the real functions $\epsilon^\Lambda (i \nu) $  and $ \gamma^\Lambda (i \nu) $ 
where $\Sigma^\Lambda(i \nu) = \epsilon^\Lambda (i \nu) + i  \gamma^\Lambda (i \nu) $.  
The initial conditions are 
\begin{align}
     \epsilon^{\Lambda \to \infty} (i \nu) = \epsilon_0-E_{\rm p} , \quad   
\gamma^{\Lambda \to \infty} (i \nu)  =0 . 
\label{eq:initialcond}     
\end{align} 
Note that not only the flow of the real and imaginary parts of $\Sigma^\Lambda$ are 
coupled but also the one of the self-energy at different frequencies [via  $\epsilon^\Lambda (\pm i \Lambda) $  
and $ \gamma^\Lambda (\pm i \Lambda) $ appearing on the right hand sides]. From the structure of the right hand sides 
and the symmetry of the initial conditions it is apparent that $\epsilon^\Lambda (i \nu)$ is even in $\nu$ 
while  $\gamma^\Lambda (i \nu)$ is odd; in the following we employ this. 

Discretizing the Matsubara 
frequency on an appropriate grid 
(which might not necessarily be equidistant; see below, in particular Appendix \ref{subsec:numflow}) 
this set of equations can easily be solved 
on a computer. When later presenting data of the numerical solution of the FRG flow equations we always 
verified that convergence with respect to the grid size as well as the lower and the upper bound 
of the grid was achieved.  

\subsection{Perturbation theory in $\lambda/\omega_0$}
\label{subsec:pt}

From FRG truncated to first order it is easy to obtain the self-energy in lowest order perturbation theory. 
For this one simply has to switch off the feedback of the self-energy on the right hand sides 
of the flow equations and replace the initial condition Eq.~(\ref{eq:initialcond}) 
for $\epsilon^{\Lambda \to \infty} (i \nu)$  by $\epsilon_0$.\cite{Metzner12} Then the
differential equations for the real and imaginary
part decouple and can be integrated leading to the Hartree and Fock parts
\begin{align} 
\Sigma_{\rm H}^{\rm pt}=  -E_p & \left[  1-\frac{2}{\pi}\arctan \left( \frac{\epsilon_0}{\Gamma} \right) 
\right] ,
\label{eq:hartree} \\
\Sigma_{\rm F}^{\rm pt}(i \nu)=  \frac{\lambda^2}{2\pi} \Bigl[ & (\tilde{d}_+-\tilde{d}_-)
\ln \left\{ -\Gamma+i \epsilon_0 \right\}  \nonumber \\
& - \tilde{d}_+\ln \left\{ \nu+i\omega_0 \right\} 
+\tilde{d}_-\ln\left\{ \nu-i\omega_0 \right\} \nonumber \\ 
& - (d_+-d_-) \ln \left\{ \Gamma+i \epsilon_0 \right\} \nonumber \\ 
 & + d_+ \ln \left\{ \nu+i\omega_0 \right\}
-d_- \ln \left\{ \nu -i\omega_0 \right\} \nonumber \\
& -i\pi (\tilde{d}_+-\tilde{d}_-) \, \mbox{sgn} \, (\epsilon_0)
-i \pi (d_++d_-) \Bigr] ,
\label{eq:fock}
\end{align} 
respectively, with  
\begin{equation} 
d_{\pm}= \frac{-1}{\epsilon_0 \pm \omega_0-i(\nu+\Gamma)}, \quad
\tilde{d}_{\pm}= \frac{-1}{\epsilon_0 \pm \omega_0-i(\nu-\Gamma)} .
\label{eq:dpmdef}
\end{equation}
These expressions can equivalently be obtained by straightforward diagrammatic perturbation theory.
The analytic continuation to the real frequency axis can be performed leading to 
\begin{align}
& \Sigma^{\rm pt,R} (\nu) = - E_{\rm p} \left[ 1 -\frac{2}{\pi}\arctan 
\left( \frac{\epsilon_0}{\Gamma} \right) \right] \nonumber \\
& +\frac{\lambda^2}{2\pi}\sum_{s=\pm} \Biggl[ i\pi \Gamma a_s \, \mbox{sgn} \, (\nu+s\omega_0)
+\Gamma a_s\ln \left\{ \epsilon_0^2+\Gamma^2 \right\} \nonumber \\
& -2 \Gamma a_s\ln \left|\nu+ s \omega_0 \right| -2 a_s \left(\nu+s\omega_0-\epsilon_0\right)
\arctan \left( \frac{\epsilon_0}{\Gamma} \right) \nonumber \\
& +\frac{\pi}{\nu-s\omega_0-\epsilon_0+i\Gamma} \Biggr]
\label{eq:fockreal}
\end{align}  
for the retarded self-energy in first order (in $\lambda^2$) perturbation theory. 
Here 
\begin{equation}
a_{\pm} =  \pm \left[ \left(\nu \pm \omega_0-\epsilon_0 \right)^2+\Gamma^2\right]^{-1} .
\label{eq:apmdef}
\end{equation}
The perturbative self-energy shows
logarithmic singularities for frequencies $\nu = \pm \omega_0$ leading to zeros in the 
spectral function (see Appendix \ref{subsec:spec-t0-comparisons}).     

To maintain the particle-hole symmetric point
it is more appropriate to consider first order perturbation theory with a propagator
dressed by a self-consistently determined Hartree self-energy. For this $\epsilon_0$ on the right hand sides 
of Eqs.~(\ref{eq:hartree})-(\ref{eq:apmdef}) must be replaced by $\epsilon_0 + \Sigma_{\rm H}$ 
and Eq.~(\ref{eq:hartree}) must be solved self-consistently. This procedure will be used in the following.

\subsection{The numerical RG}
\label{subsec:nrg}
We briefly introduce the NRG procedure applied to the SAHM which
provides an accurate description of 
physical properties in all parameter regimes of interest to us. 

The main ingredient of this approach is the logarithmic discretization of 
the reservoir(s) dispersion $\varepsilon_{k}\rightarrow \varepsilon_{\pm , n}, n=0,1,\dots$ with
$\varepsilon_{\pm , n=0}=\pm D$ and $\varepsilon_{\pm , n}= \pm D\Lambda^{-n+1-z},n=1,2,\dots$. 
It is controlled via two parameters, namely the scale parameter $\Lambda$ ($\to 1$) 
and the so-called $z$-averaging parameter $z$ which takes $N_z$ values $\in (0,1]$.
The scale parameter characterizes the relative spacing of the energy
intervals while the $z$-averaging parameter provides different realizations of 
the discretized band with the same relative spacing.
Averaging physical observables over such different realizations
largely eliminates discretization induced oscillations in physical quantities
occurring at scale parameters $\Lambda > 1$.\cite{CampoOliveira05}

The scale parameter $\Lambda$ used in NRG should not be confused with the FRG cutoff $\Lambda$. In the 
respective literature on NRG and FRG using this symbol for the two parameters is standard and we thus 
accept this double meaning.

The next crucial step is the mapping to a semi-infinite chain where the impurity is only coupled 
to the first site (representing a single conduction fermion degree of freedom). 
Following the standard tridiagonalization procedure,\cite{Wilson75,Krishnamurthy80,Bulla08}
we can find the desired chain Hamiltonian as $\mathcal{H}=\mathcal{H}_{M\to\infty}$, where
\begin{align}
\mathcal{H}_M=&H_{\rm mol} +\sqrt{\frac{\Gamma}{\pi\rho_{\rm lead}}} \sum_{\alpha=1}^{2} \left( d^\dag f_{\alpha,0} + \mbox{H.c.}  \right)\nonumber \\
&+\sum_{\alpha=1}^{2}\sum_{n=0}^{M} t^z_n \left( f^{\dagger}_{\alpha,n}f_{\alpha,n+1} + \mbox{H.c.}  \right) ,
\label{eq:semiinfchain}
\end{align}
$\{f_{\alpha,n}\}$ is a new set of mutually orthogonal (Wannier
orbital) operators constructed from linear combinations of the
original set $\{c_{\alpha,\pm n}\}$ (defining the logarithmically
discretized band) 
and $t^z_n \sim \Lambda^{-\frac{(n-1+z)}{2}}$ is the hopping amplitude from the $n$th site of the chain to the $(n+1)$th one.
An iterative diagonalization can be set up using the following
recursive formula between the truncated Hamiltonians
\begin{eqnarray}
\mathcal{H}_{M+1}= \mathcal{H}_M + t^z_n \left( f^{\dagger}_{\alpha,n}f_{\alpha,n+1} + \mbox{H.c.}  \right).
\label{eq:retruncated}
\end{eqnarray}
This can also be regarded as the RG transformation $\mathcal{T}$ such
that $\mathcal{T} [ \mathcal{H}_M ] =\mathcal{H}_{M+1}$.\cite{Krishnamurthy80} 
The maximum chain length $N$, required to describe the full spectrum of the 
Hamiltonian $\mathcal{H}$ at zero temperature, can be chosen such that
$\beta^{-1}_{N}\equiv\Lambda^{\frac{-(N-1)}{2}}\ll\Gamma \exp{\left\{-(\lambda/\omega_0)^2\right\}}$
to capture the well known polaronic suppression.\cite{Lang62,Mahan90}  

The dimension of the Hamiltonian matrix in the very first iteration, $\mathcal{H}_{M=-1}=H_{\rm mol}$, 
is already infinite due to the presence of bosonic degrees of freedom.
However, as has been established earlier,\cite{Hewson02} we can resolve the
low-energy behavior of the system with a finite number of bosons $N_{\rm b}$; for a 
given electron-phonon coupling $\lambda$ one has to keep $\gtrsim \Big(\big({\lambda}/{\omega_0}\big)^2+
5{\lambda}/{\omega_0}$\Big) phonons.\cite{Ingersent15}
Therefore, the dimension of the Hilbert space of
$\mathcal{H}_{M=-1}=H_{\rm mol}$ is $2\times N_{\rm b}$ at the first iteration and it grows by a factor of $4$ 
at each stage. Due to this exponential growth in the dimension of the Hilbert space,
we are forced to neglect high-energy states beyond some iteration $m=m_{0}$ and
retain only the first $N_{\rm s}$ ($\sim 1500$) low-energy states, thereby
keeping the calculations feasible at each iteration $m=m_{0},m_{0}+1,\dots$.

To calculate a dynamical quantity, such as the molecular dot spectral
function $A(\nu)\equiv -\frac{1}{\pi}{\rm Im}\{ G^{\rm R}_{\rm mol}(\nu)\}$, with $G^{\rm R}_{\rm mol}(\nu)=\langle \langle d;d^{\dagger}\rangle \rangle_{\nu+i\eta}
=-i\int_{0}^{\infty}dt \langle [d(t),d^{\dag}(0)]_+\rangle e^{i(\nu+i\eta)t}$, $\eta\to 0^+$
and $\langle\cdots\rangle$ denoting the thermal expectation value, we follow the procedure of Ref.~\onlinecite{CampoOliveira05}.
At vanishing temperature and a given frequency $\nu$, we choose the best shell $M$ for this frequency such that
$t^{z}_{M-1}\leq\nu<t^{z}_{M-2}$ and obtain
\begin{eqnarray}
A(\nu)= \frac{1}{\mathcal{Z}_{M}}\sum_{n,l=1}^{N_s}|\bra{n}d\ket{l}|^2 \delta(\nu-E^M_n+E^M_l)\nonumber\\
\times (e^{-\beta_{N}E^M_n}+e^{-\beta_{N}E^M_l}).
\label{eq:spectralfunc}
\end{eqnarray}
Here $\mathcal{Z}_M=\sum_{n}\exp\{-\beta_NE^M_n\}$ denotes the partition function of the best shell $M$; 
$\{\ket{n}\}$ are the eigenvectors and $\{E^M_n\}$ the eigenvalues of $\mathcal{H}_M$. We use the standard 
logarithmic Gaussian broadening with dimensionless parameter $b=0.3$.\cite{Bulla08}

The phonon contribution to the fermionic (retarded) self-energy $\Sigma^{\rm R}(\nu)$ at frequency
$\nu$ can be calculated within NRG in terms 
of the retarded Green functions $F^{\rm R}(\nu)=\langle \langle (b+b^{\dagger})d^\dagger;d\rangle \rangle_{\nu+i\eta}$ and
$G^{\rm R}_{\rm mol}(\nu)$ via $\Sigma^{\rm R}(\nu)=\frac{F^{\rm R}(\nu)}{G^{\rm R}_{\rm mol}(\nu)}$.\cite{Bulla98,Jovchev13}
Using $\Sigma^{\rm R}(\nu)$, and the exact self-energy contribution from the
reservoirs $-i\Gamma$, allows the spectral function $A(\nu)$ to be
calculated via
\begin{eqnarray}
A(\nu)=-\frac{1}{\pi}\rm Im
\left\{\frac{1}{\nu-\epsilon_{0}-\Sigma^{\rm R}(\nu) +i\Gamma}\right\}.
\label{eq:phselfspectral}
\end{eqnarray}
This approach to calculating $A(\nu)$ can significantly improve the
spectral function as compared to Eq.~(\ref{eq:spectralfunc})
, as discussed in more detail in Ref.~\onlinecite{Bulla98}.
For all the subsequent calculations, we used $\Lambda=4$, $N_z=4$ [only for Fig.~\ref{fig:spec} (a), we used $N_z=6$], $N_{\rm b}=40$ (see Appendix \ref{subsec:nb}) and $\Gamma=10^{-5}D$.
\section{Results}
\label{sec:results}

\subsection{The effective tunneling rate at particle-hole symmetry: analytical insights}
\label{subsec:Gammaeffphana}

As a first application we study the FRG flow equations at the particle-hole symmetric 
point $\epsilon_0 = E_{\rm p}$. In this case the initial condition 
Eq.~(\ref{eq:initialcond}) for the effective level position is 
$\epsilon^{\Lambda \to \infty} (i \nu) = 0$.  The flow equation
(\ref{eq:flowepsilon}) then implies $\epsilon^{\Lambda} =0$ for all $\Lambda$. The remaining
equation (\ref{eq:flowgamma}) can be simplified to
 \begin{align}
 \partial_{\Lambda}  \gamma^\Lambda (i \nu) = &     
\frac{4 \omega_0 \lambda^2}{\pi} 
\frac{ 1 }{\left|  \Lambda + \Gamma - \gamma^\Lambda(i \Lambda) \right| } \nonumber \\
& \times \frac{  \nu \Lambda}{\left[ (\nu - \Lambda)^2 + \omega_0^2 \right]
\left[ (\nu + \Lambda)^2 + \omega_0^2 \right]}  .
\label{eq:flowgammaph}
\end{align}
Defining a frequency grid this equation can easily be solved on a computer using standard 
routines. For details on this, see Appendix \ref{subsec:numflow}.  
The propagator of the molecular level at the end of the RG flow is given by ($\nu \geq 0$)  
\begin{equation}
G_{\rm mol}(i \nu) = \left[ i \nu + i \Gamma - i \gamma(i \nu) \right]^{-1} 
\label{eq:propend}
\end{equation}
where we defined $\gamma(i \nu) = \gamma^{\Lambda=0}(i \nu)$. 

To read off the renormalized tunnel coupling we Taylor expand 
\begin{equation}
\gamma^\Lambda(i \nu) = \gamma_1^\Lambda \nu + \gamma_3^\Lambda \nu^3 + \ldots 
\label{eq:gammataylor}
\end{equation}
employing that $\gamma^\Lambda(i \nu)$ is odd 
and rewrite the propagator for small $\nu$ at the end of the flow as
\begin{align}
G_{\rm mol}(i \nu) & \approx (1- \gamma_1) \left[ i \nu + i \Gamma/(1- \gamma_1)  \right]^{-1} .
\end{align}
From this expression the renormalized tunneling rate follows as
\begin{align}
\Gamma_{\rm eff} = \Gamma/(1- \gamma_1) \approx \Gamma (1 + \gamma_1) , 
\label{eq:Gammaeffex}
\end{align}
where in the last step we used that $\gamma_1$ is small if $\lambda \lessapprox \omega_0$. In fact, the
second expression to relate $\Gamma_{\rm eff}$ and $\gamma_1$ turns out to be more consistent.

Using Eqs.~(\ref{eq:flowgammaph}) and (\ref{eq:gammataylor}) we can write 
down a flow equation for the dimensionless 
first Taylor coefficient $\gamma_1^\Lambda$ of $\gamma^\Lambda (i \nu)$ 
 \begin{equation}
 \partial_{\Lambda}  \gamma_1^\Lambda  =      
\frac{4 \omega_0 \lambda^2}{\pi} 
\frac{ 1 }{\left|  \Lambda + \Gamma - \gamma^\Lambda(i \Lambda) \right| } \frac{ \Lambda}{\left( \Lambda^2 + \omega_0^2 \right)^2}
\label{eq:flowgamma1ph}
\end{equation}     
which, however, is not closed as the full function $ \gamma^\Lambda(i \nu)$ appears 
on the right hand side. The use of this equation thus requires further 
considerations. Before presenting these in Sect.~\ref{subsubsec:approx} we 
next use Eq.~(\ref{eq:flowgamma1ph}) to derive an expression for 
$\Gamma_{\rm eff}$ in lowest order perturbation theory. 

\subsubsection{Lowest order perturbation theory in $\lambda/\omega_0$}

For $\epsilon_0 = E_{\rm p}$ a self-consistent solution of the Hartree 
equation [Eq.~(\ref{eq:hartree}) with $\epsilon_0 \to \epsilon_0 + \Sigma_{\rm H}$] 
is given by $\Sigma_{\rm H}=-E_{\rm p}$. It turns out to be unique  as long as 
$E_{\rm p}/\Gamma=\lambda^2/(\omega_0 \Gamma) < \pi/2$. 
Thus $\epsilon_0+\Sigma_{\rm H}$ appearing on the right hand side of the Fock part of the self
energy computed with the Hartree propagator [Eq.~(\ref{eq:fock}) with $\epsilon_0 \to 
\epsilon_0 + \Sigma_{\rm H}$] vanishes in this case and Eq.~(\ref{eq:flowgamma1ph}) with the self-energy 
feedback set to zero provides an equation for $\gamma_1$ to lowest order in $\lambda^2$. 
 It reads
  \begin{equation}
 \partial_{\Lambda}  \left(\gamma_1^{\rm pt}\right)^\Lambda  =      
\frac{4 \omega_0 \lambda^2}{\pi} 
\frac{ 1 }{ \Lambda + \Gamma }  \frac{ \Lambda}{\left( \Lambda^2 + \omega_0^2 \right)^2}   
\label{eq:flowgamma1phpt}
\end{equation}    
and can be integrated from $\Lambda=\infty$ down to $\Lambda=0$ employing the initial condition
$ \left(\gamma_1^{\rm pt}\right)^{\Lambda=\infty}=0$. Inserting $ \gamma_1^{\rm pt} = 
\left(\gamma_1^{\rm pt}\right)^{\Lambda=0}$
into the second relation of Eq.~(\ref{eq:Gammaeffex}) we obtain in lowest order perturbation theory
   \begin{align}
\frac{\Gamma_{\rm eff}^{\rm pt}}{\Gamma}  & =   1  
- \left( \frac{\lambda}{\omega_0}\right)^2 \left[1+ \left( \frac{\Gamma}{\omega_0}\right)^2  
\right]^{-2} \nonumber \\
 & \times \left[ 1 + \frac{4}{\pi} \frac{\Gamma}{\omega_0} \ln \frac{\Gamma}{\omega_0} + 
  \frac{\Gamma}{\omega_0} \left\{ \frac{2}{\pi} -  \frac{\Gamma}{\omega_0} +   \frac{2}{\pi} 
 \left( \frac{\Gamma}{\omega_0}\right)^2   \right\}\right] .
\label{eq:Gammaeffpt}
\end{align}   

In the {\it adiabatic limit} $\Gamma \gg \omega_0$ this reduces to the well known 
result\cite{Vinkler12,Jovchev13}
\begin{align}
\frac{\Gamma_{\rm eff}^{\rm pt}}{\Gamma}  & = 1- \frac{2}{\pi} \left( \frac{\lambda}{\omega_0}\right)^2 
\frac{\omega_0}{\Gamma} + {\mathcal O}\left([\omega_0/\Gamma]^2\right) \nonumber \\ & = 
1- \frac{2}{\pi} \frac{E_{\rm p}}{\Gamma} + {\mathcal O}\left([\omega_0/\Gamma]^2\right) . 
\label{eq:Gammaeffptadi}
\end{align}
In the {\it antiadiabatic regime} $\Gamma \ll \omega_0$ we obtain
\begin{align}
\frac{\Gamma_{\rm eff}^{\rm pt}}{\Gamma}   = 1 -  \left( \frac{\lambda}{\omega_0}\right)^2 
\Biggl[ & 1 + \frac{4}{\pi} \frac{\Gamma}{\omega_0} \ln \frac{\Gamma}{\omega_0} + 
 \frac{2}{\pi}   \frac{\Gamma}{\omega_0} \nonumber \\ &+ 
{\mathcal O}\left(\{\Gamma/\omega_0\}^2\right) \Biggr] .
\label{eq:Gammaeffptantiadi}
\end{align}

This result should be compared to the lowest order Taylor expansion in $\lambda/\omega_0$ of 
Eq.~(\ref{eq:Gammaeff}) obtained by the mapping to an IRLM
\begin{align}
\frac{\Gamma_{\rm eff}^{\rm IRLM}}{\Gamma}   = 1 -  \left( \frac{\lambda}{\omega_0}\right)^2 
\Biggl[ & 1 + \frac{4}{\pi} \frac{\Gamma}{\omega_0} \ln \frac{\Gamma}{\omega_0} 
 \nonumber \\ & + {\mathcal O}\left( \left\{  \frac{\Gamma}{\omega_0} \ln \frac{\Gamma}{\omega_0}  
\right\}^2\right) \Biggr] .
\label{eq:Gammaeffantiadi}
\end{align}
This shows that the mapping only holds up to order $\frac{\Gamma}{\omega_0} 
\ln \frac{\Gamma}{\omega_0}$ (at least for small $\lambda/\omega_0$); already the 
linear term $\sim \Gamma/\omega_0$ is not properly represented. This
defines the limit of the mapping of the SAHM to the IRLM in the antiadiabatic 
regime.   

\subsubsection{Approximate solution of the FRG equation}
\label{subsubsec:approx}

When numerically integrating  the full lowest order flow equation (\ref{eq:flowgammaph}) 
from  $\Lambda=\infty$ to $\Lambda=0$, $\gamma^\Lambda(i \Lambda)$ takes sizable values [starting 
at $\gamma^{\Lambda=\infty}(i \nu)=0$ for all $\nu$] only when $\Lambda$ is so small 
that one can linearize $\gamma^\Lambda(i \Lambda) \approx \gamma^\Lambda_1 \Lambda$. Inserting this
expansion on the right hand side of Eq.~(\ref{eq:flowgamma1ph}) leads to a closed equation 
for $\gamma^\Lambda_1$
 \begin{equation}
 \partial_{\Lambda}  \gamma_1^\Lambda  \approx      
\frac{4 \omega_0 \lambda^2}{\pi} \frac{1}{1-\gamma^\Lambda_1} 
\frac{1}{\Lambda+\frac{\Gamma}{1-\gamma^\Lambda_1}}
\frac{\Lambda}{\left(\Lambda^2+\omega_0^2 \right)^2} .
\label{eq:flowgamma1phlin}
\end{equation}  
As $\gamma_1^\Lambda \sim \lambda^2$ we can further expand
 \begin{equation}
 \partial_{\Lambda}  \gamma_1^\Lambda  \approx      
\frac{4 \omega_0 \lambda^2}{\pi} \left( 1 + \gamma^\Lambda_1\right) 
\frac{1}{\Lambda+\Gamma \left(1+\gamma^\Lambda_1\right)}
\frac{\Lambda}{\left(\Lambda^2+\omega_0^2 \right)^2} .
\label{eq:flowgamma1phlin1}
\end{equation}    

For $\Gamma \ll \omega_0$ the last factor of Eq.~(\ref{eq:flowgamma1phlin1}) ensures that $\gamma_1^\Lambda$ changes 
significantly only on the scale $\Lambda \approx \omega_0$. In this regime 
$\Gamma \gamma_1^\Lambda$ in the denominator of the second to last factor 
can be neglected as compared to $\Lambda + \Gamma$.  
In this antiadiabatic regime the differential flow equation thus reduces to
 \begin{equation}
\frac{ \partial_{\Lambda} \left( 1 + \gamma_1^\Lambda \right)}{1+\gamma_1^\Lambda} =  
\frac{4 \omega_0 \lambda^2}{\pi}
\frac{1}{\Lambda+\Gamma}
\frac{\Lambda}{\left(\Lambda^2+\omega_0^2 \right)^2} ,  
\label{eq:flowgamma1phlin1anti}
\end{equation} 
which according to Eq.~(\ref{eq:Gammaeffex}) is a flow equation for $\Gamma_{\rm eff}/\Gamma$.  
Remarkably the right hand side has exactly the form as obtained in perturbation 
theory Eq.~(\ref{eq:flowgamma1phpt}). Rewriting Eq.~(\ref{eq:flowgamma1phlin1})
as 
\begin{equation}
 \partial_{\Lambda}  \gamma_1^\Lambda  =       
\frac{4 \omega_0 \lambda^2}{\pi} \left( 1 + \gamma^\Lambda_1\right) 
\frac{1}{\Lambda+\Gamma} \frac{1}{1+ \frac{\Gamma \gamma^\Lambda_1}{\Lambda+\Gamma}}
\frac{\Lambda}{\left(\Lambda^2+\omega_0^2 \right)^2} 
\label{eq:flowgamma1phlin1adi}
\end{equation}     
it is obvious that in the adiabatic limit $\Gamma \gamma^\Lambda_1/(\Lambda+\Gamma)$ can be neglected 
as compared to 1 and we recover Eq.~(\ref{eq:flowgamma1phlin1anti}); it is thus tempting to 
conclude that Eq.~(\ref{eq:flowgamma1phlin1anti}) is valid for all $\Gamma/\omega_0$. 
The solution of this equation is given by 
\begin{align}
& \frac{\Gamma_{\rm eff}^{\rm FRG}}{\Gamma}  =  \exp \Biggl\{   
- \left( \frac{\lambda}{\omega_0}\right)^2 \left[1+ \left( \frac{\Gamma}{\omega_0}\right)^2  
\right]^{-2} \nonumber \\
 & \times \left[ 1 + \frac{4}{\pi} \frac{\Gamma}{\omega_0} \ln \frac{\Gamma}{\omega_0} + 
  \frac{\Gamma}{\omega_0} \left\{ \frac{2}{\pi} -  \frac{\Gamma}{\omega_0} +   \frac{2}{\pi} 
 \left( \frac{\Gamma}{\omega_0}\right)^2   \right\}\right] \Biggr\}.
\label{eq:Gammaeffana}  
\end{align}
Figure \ref{fig:geffall} shows that for sufficiently small $\lambda$, in which our lowest order 
FRG approach is controlled Eq.~(\ref{eq:Gammaeffana}) agrees rather 
well with the $\Gamma_{\rm eff}$ obtained from the numerical solution of the full lowest order 
flow equation (\ref {eq:flowgammaph}) as well as with the renormalized tunneling rate 
computed using NRG for all $\Gamma/\omega_0$; for more, see the next section.   

In the adiabatic
regime $\Gamma \gg \omega_0$ Eq.~(\ref{eq:Gammaeffana}) reduces to the perturbative
result Eq.~(\ref{eq:Gammaeffptadi}).  In the opposite antiadiabatic limit 
$\Gamma \ll \omega_0$ in which only the term $ \frac{\Gamma}{\omega_0} \ln 
\frac{\Gamma}{\omega_0}$ in the argument of the exponential 
function in Eq.~(\ref{eq:Gammaeffana}) is kept we exactly reproduce the small $\lambda$ 
result for $\Gamma_{\rm eff}$ obtained by the mapping to the IRLM Eq.~(\ref{eq:Gammaeff}).
The lowest order truncated FRG thus provides a proper resummation of diagrams 
(perturbative in $\lambda$) to reproduce the involved interplay of exponential 
(polaronic) as well as power-law (x-ray edge) renormalization in the electron-phonon 
coupling $\lambda$. 
From the perspective of the method this remarkable result provides another example that 
essentially analytical truncated FRG, which leads to transparent equations, can 
be used to study complex many-body physics including correlation 
effects.\cite{Metzner12,Karrasch10,Karrasch06} 
From the perspective of the physics Eq.~(\ref{eq:Gammaeffana}) provides a remarkably
simple closed expression for the renormalized tunneling rate at small to intermediate 
electron-phonon coupling $\lambda$ going way beyond 
lowest order perturbation theory in $\lambda$ which is 
(approximately) valid for all $\Gamma/\omega_0$.

\subsection{Numerical results for the effective tunneling rate at particle-hole symmetry}
\label{subsec:Gammaeffphnum}
We want to verify the validity of the approximated tunneling rate 
Eq.~(\ref{eq:Gammaeffana}) obtained analytically in the previous section, by comparing 
it to the numerical solution of the full first order truncated FRG flow equation 
(\ref{eq:flowgammaph}) and also to NRG results. In NRG, the renormalized 
tunneling rate, is calculated from the charge susceptibility:
\begin{equation}
\Gamma_{\rm eff}=\frac{1}{\pi\chi_{\text{c}}},
\label{eq:geff}
\end{equation}
with
\begin{equation}
\chi_{\text{c}}=-\frac{dn_{\rm mol}(\epsilon_0)}{d\epsilon_0}\bigg|_{\epsilon_0=E_{\rm p}}.
\end{equation} 
The occupancy $n_{\rm mol}$ of the molecular level can be calculated as, 
\begin{eqnarray}
n_{\rm mol}=\frac{1}{\mathcal{Z}_N}\sum_{n}\bra{n}d^{\dagger}d\ket{n}e^{-\beta_NE^N_n},
\end{eqnarray}
with $\{E^N_n\}$ and $\{\ket{n}\}$ being the eigenvalues and the eigenvectors, of the longest chain Hamiltonian $\mathcal{H}_N$.

As mentioned before, the RG resummation within lowest order truncated FRG is well controlled up to first 
order in $\lambda^2$. After numerically solving the full flow equation 
(\ref{eq:flowgammaph}) for $\gamma(\nu)$ for consistency we thus compute the effective tunneling rate as 
in Eq.~(\ref{eq:Gammaeffex}) by expanding
$\Gamma_{\rm eff}/\Gamma \approx 1+\frac{d\gamma(i\nu)}{d\nu}|_{\nu=0}$. 
The details of the numerical implementation of the solution of the flow equation can be found 
in Appendix \ref{subsec:numflow}.

Figure \ref{fig:geffall} (a) shows a comparison of $\Gamma_{\rm eff}$ obtained by the different methods 
introduced for electron-phonon coupling $\lambda=0.5\omega_0$ all the way from the antiadiabatic limit to 
the adiabatic one (note the logarithmic x-axis scale).
In the adiabatic limit the results of all the methods agree very well, however, 
as we approach the antiadiabatic limit, the purely perturbative result Eq.~(\ref{eq:Gammaeffpt}) starts 
to deviate. In particular, it fails to produce the result obtained from the mapping to the 
IRLM Eq.~(\ref{eq:Gammaeff}); in the antiadiabatic regime the physics is nonperturbative even at fairly small 
$\lambda/\omega_0$. The nice match of the NRG and the FRG data sets proves that this physics can indeed be captured within 
lowest order truncated FRG. Additionally, the agreement of the analytical expression Eq.~(\ref{eq:Gammaeffana}) 
to the result from the numerical solution of the full flow Eq.~(\ref{eq:flowgammaph})  
['FRG' in Fig.~\ref{fig:geffall} (a)] verifies the validity of the approximations introduced in 
Sec.~\ref{subsubsec:approx}.
\begin{figure}[!htbp]
 \centering
 \graphicspath{ {plots/}}
 \includegraphics[width=0.9\linewidth]{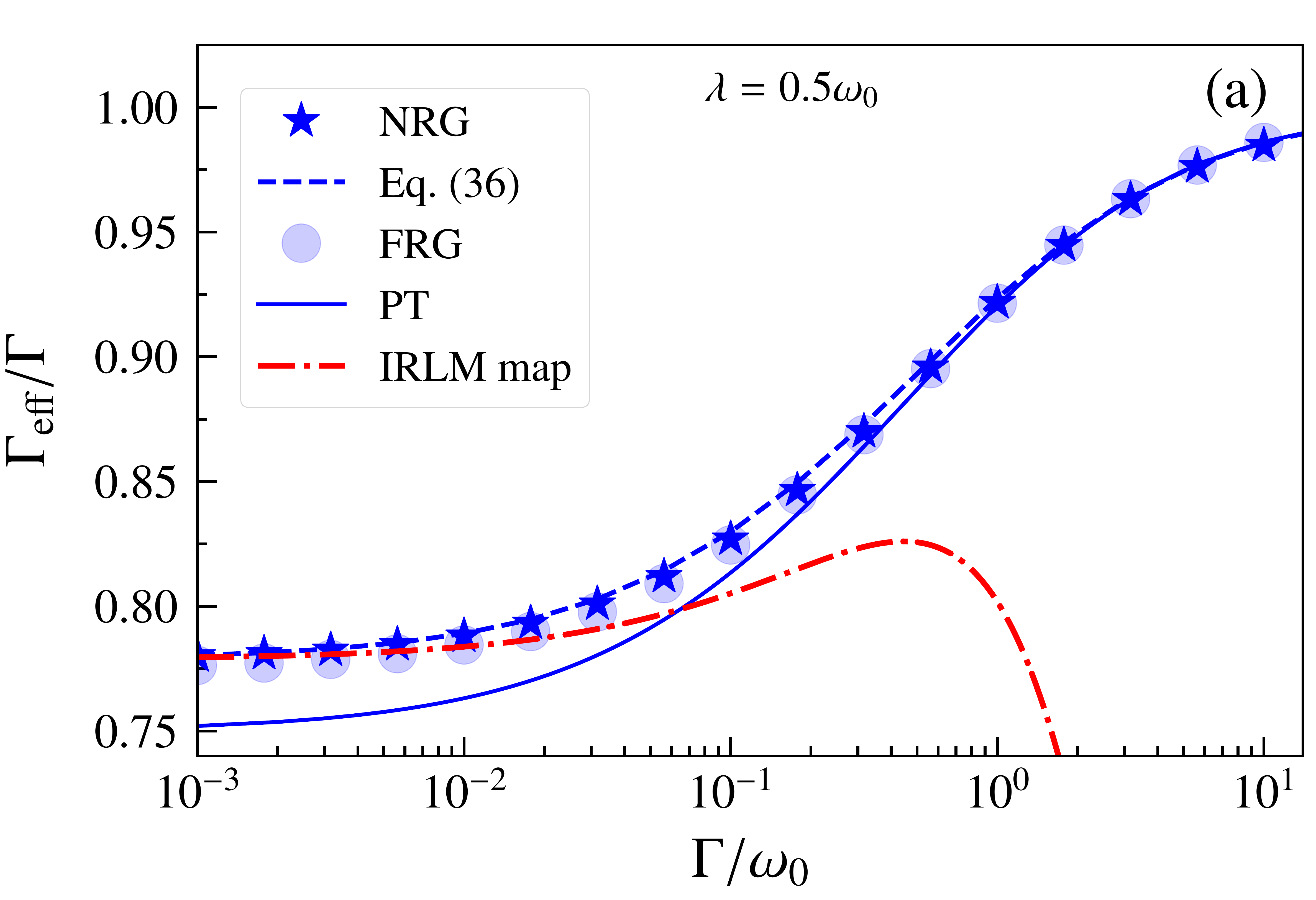}
  \includegraphics[width=0.9\linewidth]{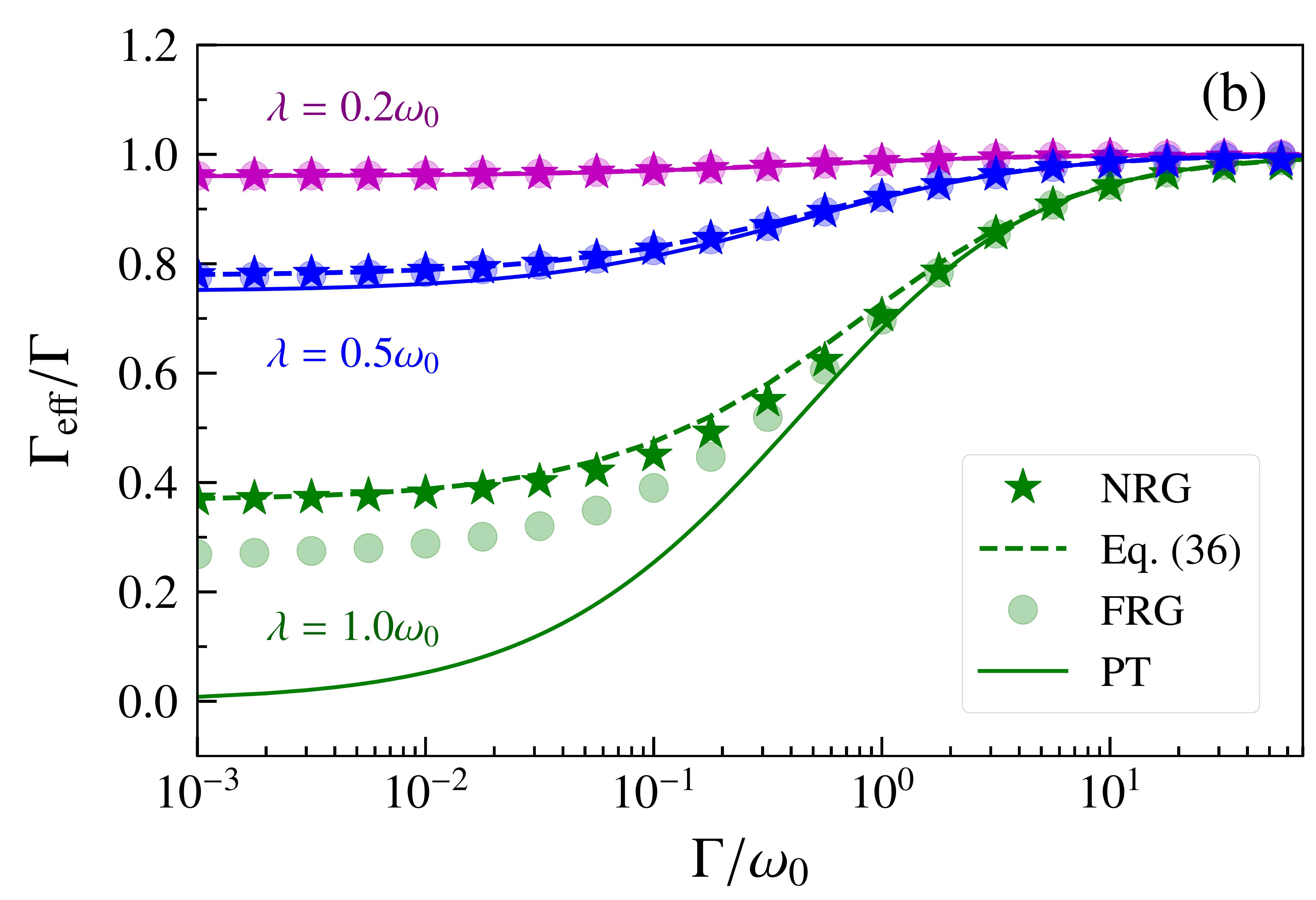}
  \includegraphics[width=0.9\linewidth]{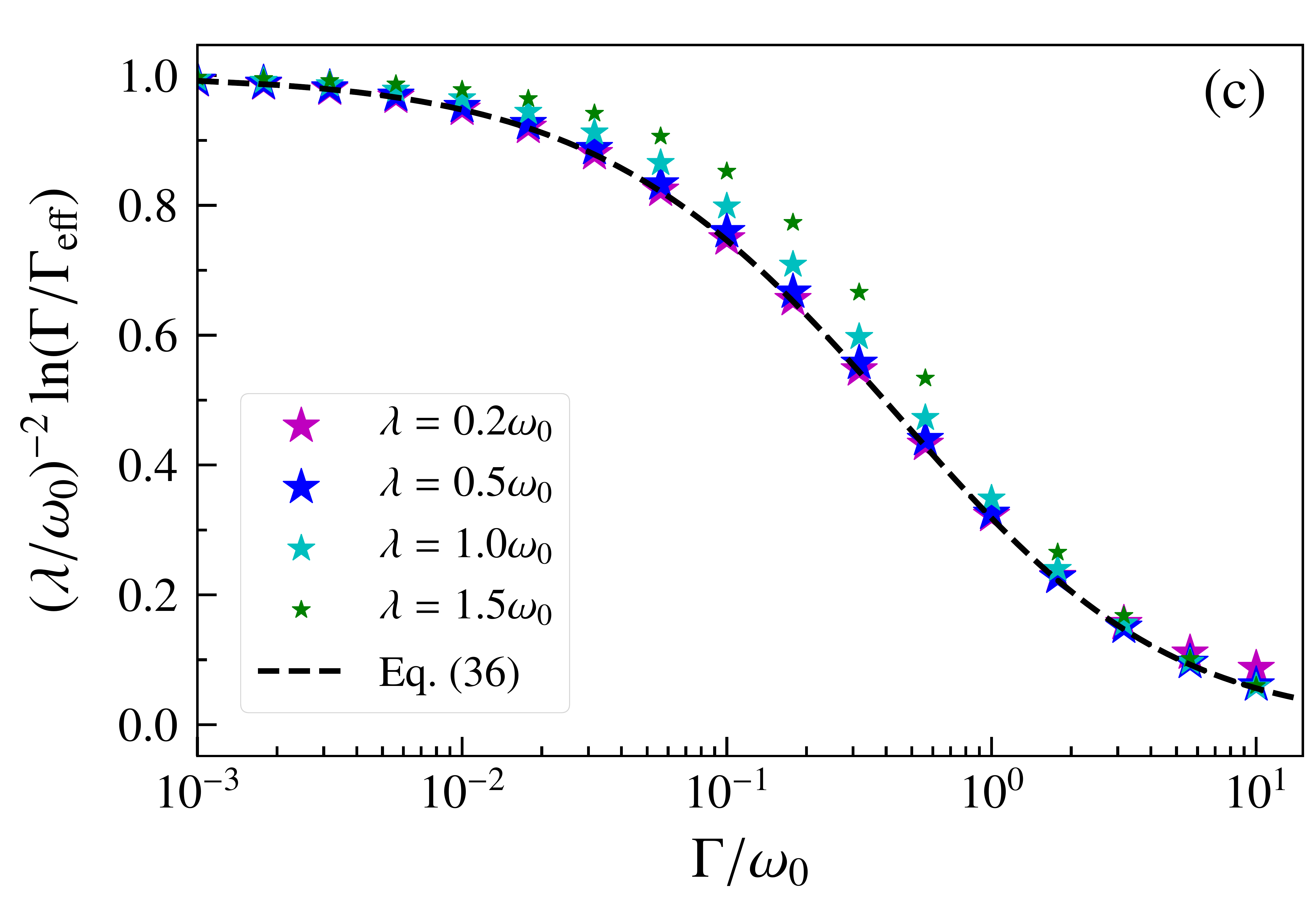}
 \caption{(a) and (b): The ratio of the effective tunneling rate $\Gamma_{\rm eff}$ to the bare value $\Gamma$ as a function of 
${\Gamma}/{\omega_0}$ using different approaches. (c) 
The ratio $\big({\lambda}/{\omega_0}\big)^{-2}\ln\Big({\Gamma_{\rm eff}}/{\Gamma}\Big)$ as
a function of $\Gamma/\omega_0$ for different electron-phonon couplings computed within NRG.}
 \label{fig:geffall}
\end{figure}

The phonon-assisted suppression of tunneling processes depends on the strength of the electron-phonon coupling, 
as it is shown in Fig.~\ref{fig:geffall} (b).
As we approach the strong coupling regime $\lambda>\omega_0$, higher order coefficients in the Taylor expansion of 
the self-energy feedback in Eq.~(\ref{eq:gammataylor}) produce sizable contributions.
Therefore the solution of the first order truncated FRG Eq.~(\ref{eq:flowgammaph}) becomes different from
the approximated formula Eq.~(\ref{eq:Gammaeffana}), which was obtained by including the linear coefficient $\gamma_1$ only.
The approximated formula matches better with the NRG data as compared to the results obtained from the numerical solution of the 
full lowest order truncated FRG for $\lambda \approx \omega_0$ which, however, must be regarded as accidental.  

Figure \ref{fig:geffall} (c) shows that also the NRG data for $\Gamma_{\rm eff}$ approximately follow a scaling 
form as it is exactly fulfilled in the approximate (in $\lambda^2$) expression Eq.~(\ref{eq:Gammaeffana}); 
$\big(\frac{\lambda}{\omega_0}\big)^{-2}\ln\Big(\frac{\Gamma_{\rm eff}^{\rm FRG}}{\Gamma}\Big)$ is only a function 
of $\frac{\omega_0}{\Gamma}$. This seems to be valid even for $\lambda \gtrapprox \omega_0$ as long as we stay 
away from the crossover regime between antiadiabatic and adiabatic; in this the exact 
solution and thus the highly accurate NRG approximation to the latter contains $\lambda/\omega_0$ dependent corrections 
to the simple scaling form. This insight is consistent to the 
earlier study,\cite{Eidelstein13} where it was found that in the strong coupling regime the 
aforementioned ratio is a function of ${E_{\rm p}}/{\Gamma}$.

\begin{figure}[h]
 \centering
 \graphicspath{ {plots/}}
 \includegraphics[width=0.9\linewidth]{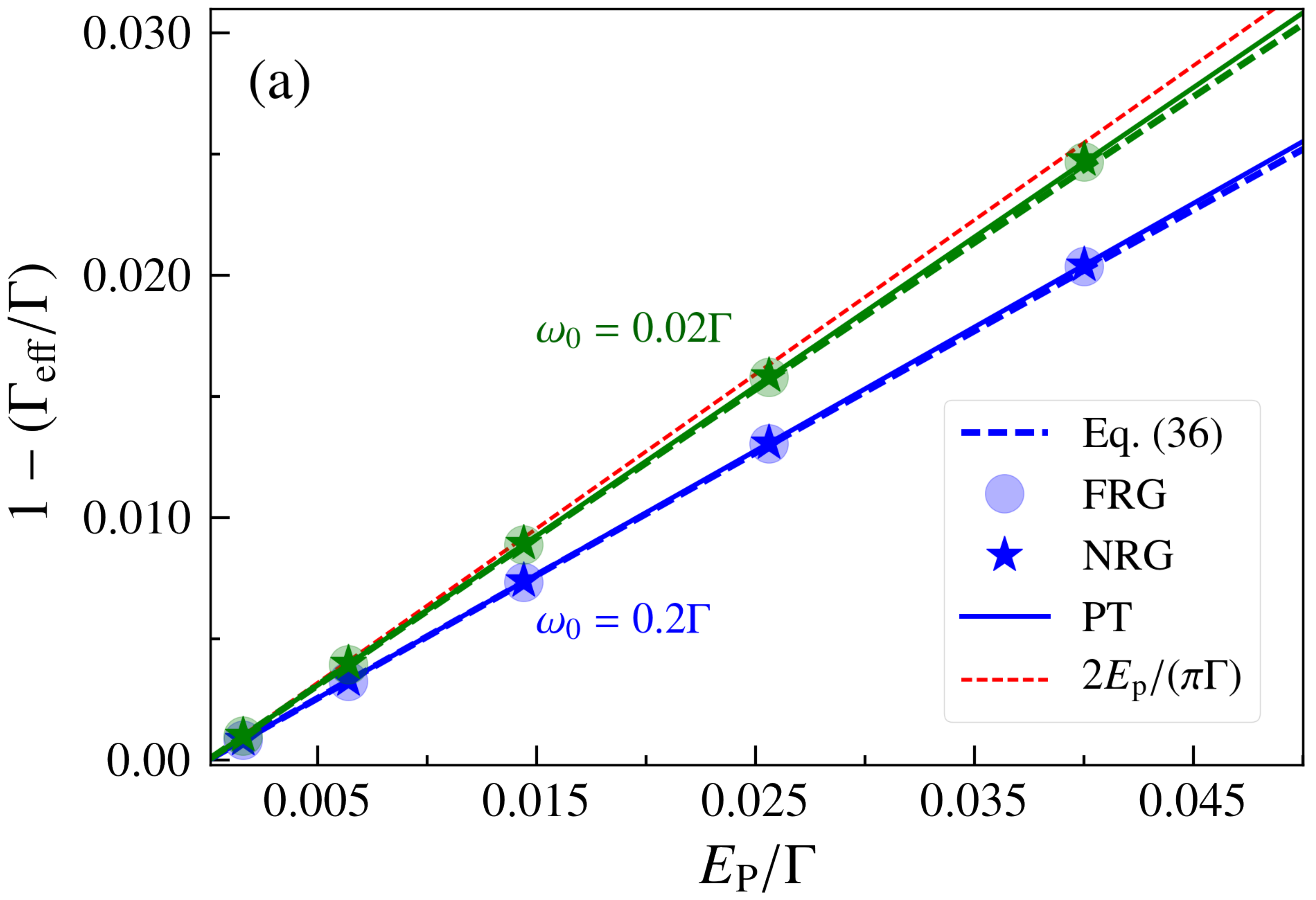}
 \includegraphics[width=0.9\linewidth]{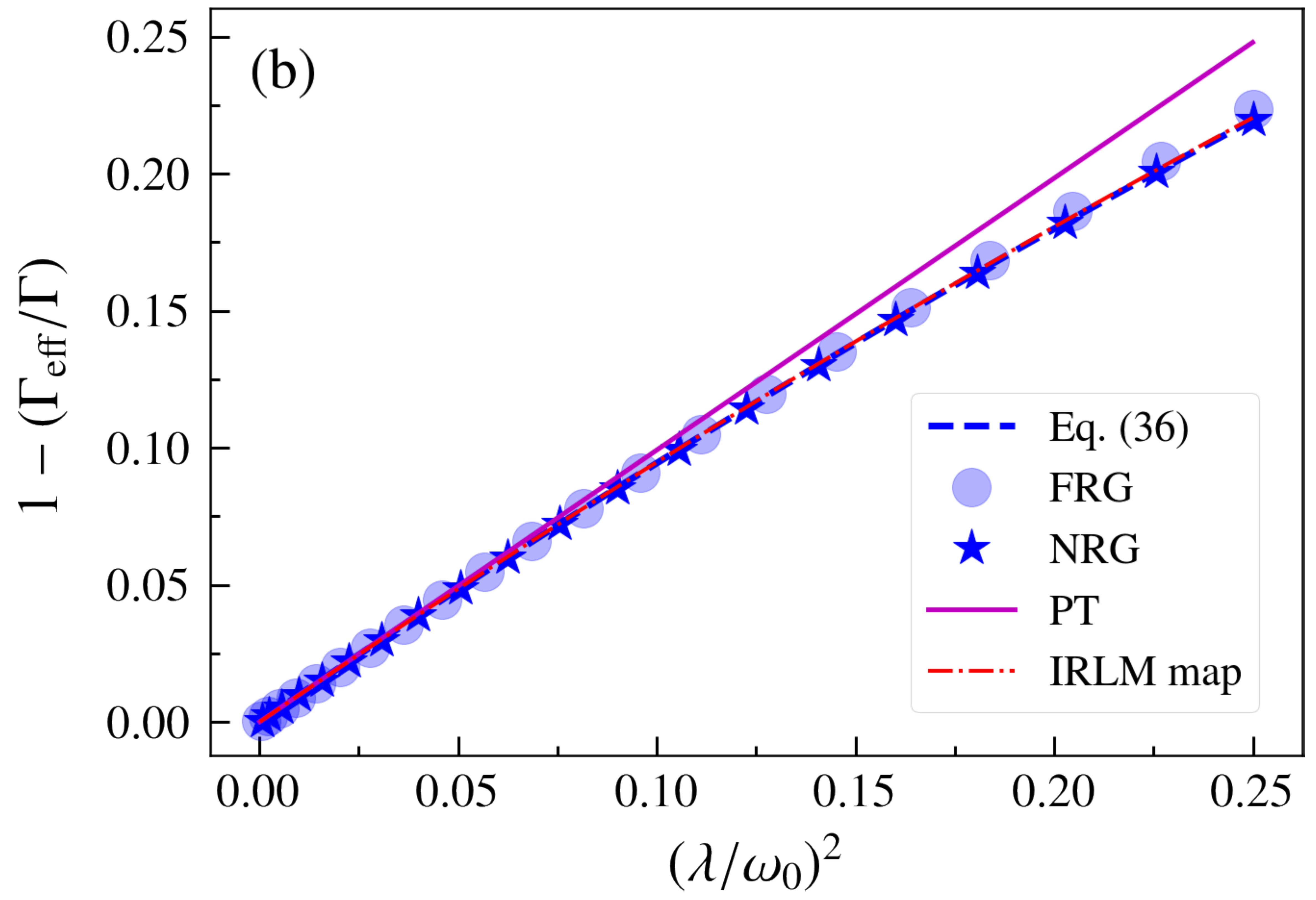}
 \caption{ (a) $1-({\Gamma_{\rm eff}}/{\Gamma})$ as a function of ${E_{\rm p}}/{\Gamma}$ 
for $\omega_0=0.2\Gamma$ and $\omega_0=0.02\Gamma$ (adiabatic limit).
(b)  $1-({\Gamma_{\rm eff}}/{\Gamma})$ as a function of 
$\big({\lambda}/{\omega_0}\big)^2$ for $\omega_0=10^3\Gamma$ (antiadiabatic limit).}
 \label{fig:geffsep}
\end{figure}

We address the two extreme antiadiabatic and adiabatic regimes separately in Fig.~\ref{fig:geffsep}.
In the adiabatic limit, the slow molecular vibrations cannot change the charge fluctuations significantly; 
conventional perturbation theory suffices and matches well with other methods [see Fig.~\ref{fig:geffsep} (a)].
As we go sufficiently deep into the adiabatic regime, we obtain the well known asymptotic behavior 
Eq.~(\ref{eq:Gammaeffptadi}). We note in passing that in Ref.~\onlinecite{Jovchev13} the ratio $\Gamma/\omega_0$ 
was not chosen large enough to properly reproduce the simple expression Eq.~(\ref{eq:Gammaeffptadi}); for the value
considered in this work corrections in $\omega_0/\Gamma$ as given in Eq.~(\ref{eq:Gammaeffpt}) must be kept.   

Figure ~\ref{fig:geffsep} (b) shows that in the antiadiabatic limit, perturbative results start to 
deviate already for small electron-phonon couplings. 

The overall physical picture is the following; phonons induce a retarded and attractive fermion-fermion
interaction on the dot [see Eq. (\ref{eq:interact})] and therefore, the tunneling 
processes from the dot into the leads are suppressed.
This suppression is more significant in the antiadiabatic regime, where phonons are quite fast 
compared to the tunneling processes. 
\subsection{Gate voltage dependence of the $T=0$ elelctrical conductance}
\label{subsec:awayph}
Having investigated within perturbation theory, FRG and NRG the emergent 
low-energy scale $\Gamma_{\rm eff}$ at particle-hole symmetry, we now
turn to the case of finite particle-hole asymmetry and investigate how
this scale manifests itself in the gate voltage dependence of the $T=0$
electrical conductance, making comparisons between the different
approaches. 

Within the FRG approach, as formulated here in Matsubara space, one
would need to analytically continue the molecular dot Green function to the real axis in order to
calculate the linear conductance from the molecular spectral function\cite{Meir92} as it is often done. This
analytic continuation is an ill-posed problem. We can, however, compute the linear conductance as a function of the (bare) level position from the FRG self-energy 
data without performing the analytic continuation by employing a continued fraction (CF) representation of the 
Fermi function $f$
\begin{equation}
f(\beta \nu)=\frac{1}{2}-\frac{1}{\beta}\sum_{p=1}^{M} \left\{ 
\frac{R_{p}}{\nu-i\frac{z_{p}}{\beta}}+
\frac{R_{p}}{\nu+i\frac{z_{p}}{\beta}} \right\} 
\label{continuedfractionFermi}
\end{equation}
at inverse temperature $\beta$. 
Here the $M$ poles $\frac{z_{p}}{\beta}$ and residues $R_{p}$ can be calculated as proposed in Refs.~\onlinecite{Ozaki07,KarraschOzaki10}.
The poles are concentrated densely close to the real axis and they are further apart as we go up and down the imaginary axis.
This leads to a very fast convergence of the sum in Eq.~(\ref{continuedfractionFermi}) as compared to the Matsubara 
representation. To obtain the conductance at vanishing temperature 
we choose a sufficiently small $\beta^{-1}=10^{-4} \Gamma^{\rm FRG}_{\rm eff}$.
Employing the above mentioned relation between the molecular spectral function $A(\nu)$ 
and the linear conductance $G$\cite{Meir92} we obtain
\begin{align}
\frac{G}{G_0}&=-\pi\Gamma\int_{-\infty}^{\infty} d\nu A(\nu)  \partial_{\nu}f(\beta\nu) \nonumber\\
&=\frac{2\pi\Gamma}{\beta}\sum_{\rm p=1}^{M} R_p \mbox{Im} \left[\frac{dG_{\rm mol}(i\frac{z_{p}}{\beta})}{d(\frac{z_{p}}{\beta})}\right] ,
\label{eq:Meir}
\end{align}
where $G_0=e^2/h$, with $e$ and $h$ denoting electric charge and Plank's constant respectively.

\begin{figure}[h]
 \centering
 \graphicspath{ {plots/}}
 \includegraphics[width=0.9\linewidth]{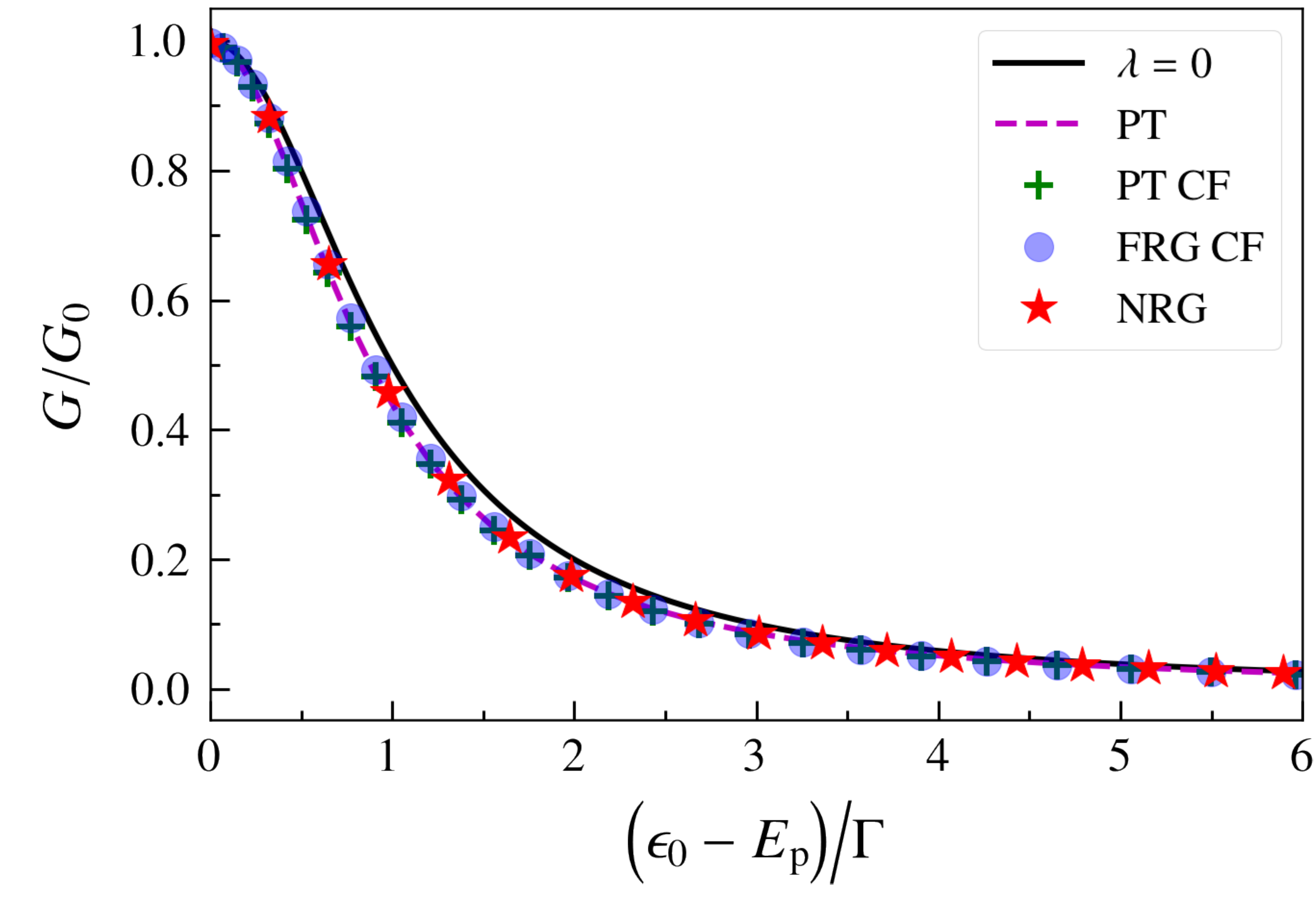}
 \caption{Comparison of the different approaches to compute the $T=0$ linear 
conductance as a function of the gate voltage 
 $\epsilon_0-E_{\rm p}$ for a given phonon frequency 
$\omega_0=2\Gamma$ in the weak coupling regime ($\lambda=0.5\omega_0$).}
 \label{fig:K1}
\end{figure}

To test the CF approach in the present context in Fig.~\ref{fig:K1} we compare the linear conductance obtained from
the self-energy in lowest order perturbation theory after performing the analytic continuation as in 
Eq.~(\ref{eq:fockreal}) [using the first line of Eq.~(\ref{eq:Meir})] (dashed line) and the continued fraction 
representation employing Matsubara frequency data  as in Eqs.~(\ref{eq:hartree}) and (\ref{eq:fock}) 
[using the last line of Eq.~(\ref{eq:Meir})] (plus signs). Both conductance curves agree well.
For the weak interaction of this figure, and on the scale of the plot, perturbation theory matches the
FRG data (obtained from the continued fraction representation; circles) as well as the NRG ones (stars). 
The latter were obtained from the spectral weight $A(\nu=0)$ [see the $T=0$ limit of the first line of Eq.~(\ref{eq:Meir})]. 
Due to the presence of phonons, the linear conductance is narrower as compared to the noninteracting case.
This effect can be captured within perturbation theory as long as ${E_{\rm p}}/{\Gamma}<{\pi}/{2}$.
Therefore, if we are deep in the antiadiabatic limit, perturbation theory is limited to extremely small 
coupling constants $\lambda/\omega_0$.   

\begin{figure}[t]
 \centering
 \graphicspath{ {plots/}}
 \includegraphics[width=0.9\linewidth]{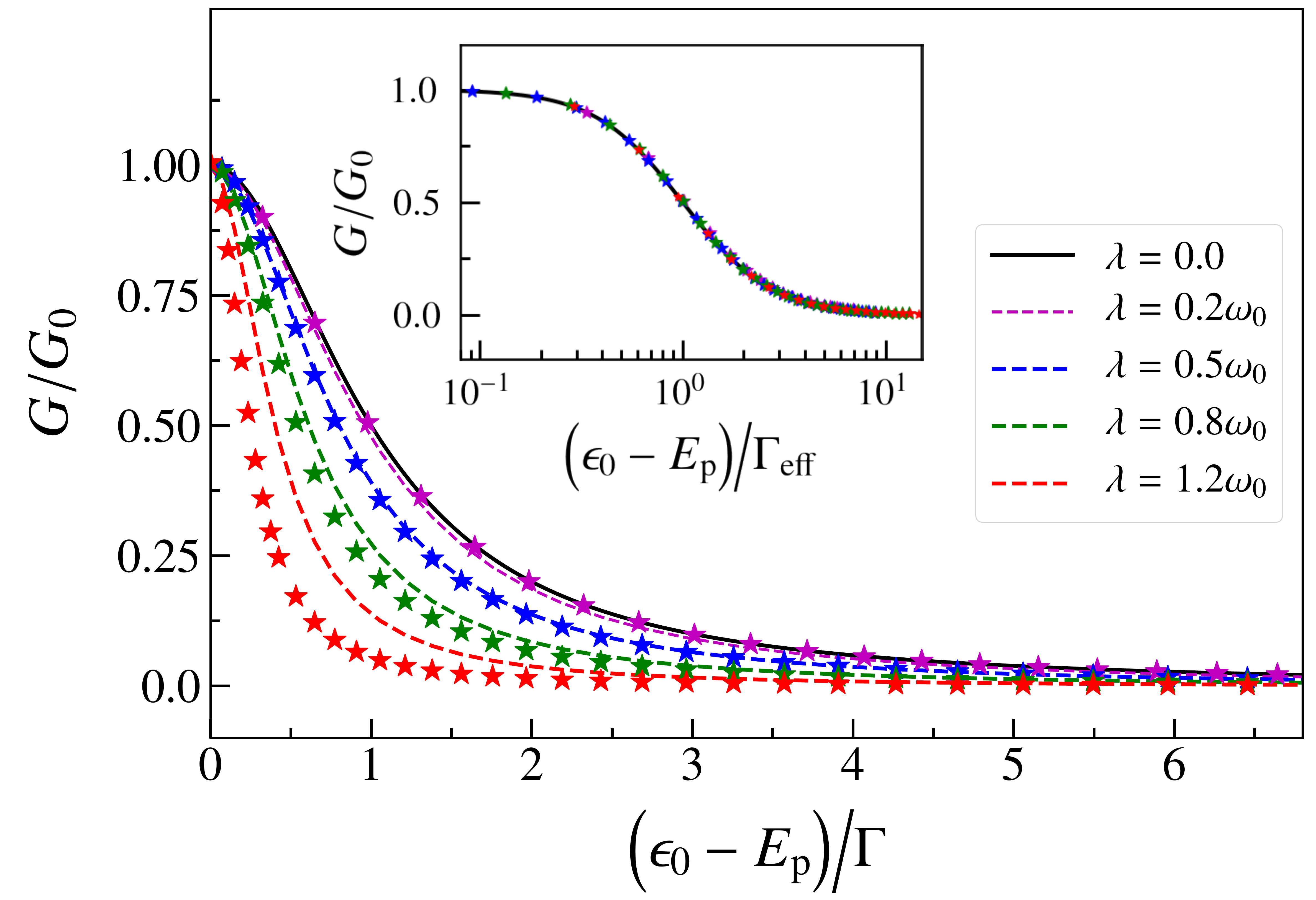}
 \caption{Comparison of FRG (dashed lines) and NRG (stars) results for the 
zero temperature linear conductance as a function of the gate voltage 
$\epsilon_0-E_{\rm p}$
 for different strength of electron-phonon couplings in the 
antiadiabatic limit ($\omega_0=10^3\Gamma$).
 The inset shows the zero temperature conductance from NRG as a function of the rescaled gate voltage $({\epsilon_0-E_{\rm p}})/{\Gamma_{\rm eff}}$ for different 
strength of the electron-phonon coupling 
in the antiadiabatic limit ($\omega_0=10^3\Gamma$).}
 \label{fig:K2all}
\end{figure}

Figure \ref{fig:K2all} shows that FRG results for $G$ match very well to the NRG data for $\lambda/\omega_0 \lessapprox 1$.
The narrowing of the linear conductance reflects that due to molecular vibrations, small misalignment 
of the gate voltage to the chemical potential of the leads can result in a substantial drop in transport. 
This is just another indication of the suppression of tunneling processes.
In fact, if we rescale the level position with the renormalized tunneling rate $\Gamma_{\rm eff}$,
all the NRG curves corresponding to different strengths of
electron-phonon coupling collapse, with good accuracy, to the noninteracting curve 
as shown in the inset of Fig.~\ref{fig:K2all}. This shows that $\Gamma_{\rm eff}$ is the relevant low-energy 
scale also away from particle-hole symmetry; the width of the linear 
conductance resonance as a function of the level position is 
given by $\Gamma_{\rm eff}$. 
\section{Summary and outlook}
\label{sec:summary}

Using a combination of lowest order perturbation theory, (truncated) FRG, and NRG we studied 
phonon assisted tunneling in an elementary model of a molecular electronics device. 
Complementing an earlier strong coupling study\cite{Eidelstein13} our focus was on weak to intermediate
electron-phonon coupling. We derived an analytic expression for the renormalized tunnel coupling 
at particle-hole symmetry valid for all ratios $\Gamma/\omega_0$ from the antiadiabatic into the 
adiabatic regime. It captures the combined exponential (polaronic) and power-law (x-ray edge singularity) 
renormalization in the antiadiabatic limit known from the mapping to an effective IRLM. 
Away from particle-hole symmetry we investigated the influence of the
emergent low-energy scale $\Gamma_{\rm eff}$ on the $T=0$ 
electrical conductance, comparing also the results within
different approaches. 

In a follow-up paper,\cite{Khedri17b} we consider a (small) {\it temperature} bias as the driving force 
and in this way extend our study to linear {\it thermoelectric properties of molecular 
devices.} Such devices are considered to be promising building blocks for waste heat conversion 
and cooling on the molecular level.\cite{Mahan97,Giazotto06} Indeed,
in Ref.~\onlinecite{Khedri17b}, employing the NRG,\cite{Costi10} we
find parameter regimes where the {\it linear}  thermoelectric
response through such a device is significantly enhanced.

In the near future we plan to further extend the FRG to the Keldysh 
contour\cite{Jakobs07,Karrasch10,Laakso14} in order to compute the equilibrium spectral function 
without the need for an analytic continuation. This will in addition enable us to
investigate the nonlinear (finite voltage and temperature bias)
thermoelectric transport properties of molecular 
devices described by the SAHM.

\begin{acknowledgements}
 
This work was supported by the Deutscheforschungsgemeinschaft via RTG 1995. We acknowledge
useful discussions with Dante Kennes in the early stages of this work and supercomputing support
by the John von Neumann Institute for Computing (J\"ulich).

\end{acknowledgements}

\appendix
\section{Numerical implementation of the flow equations}
\label{subsec:numflow}
To solve the coupled differential Eqs.~(\ref{eq:flowepsilon}) and (\ref{eq:flowgamma}), first we discretize the Mathsubara frequency.
In the light of the emergent low-energy scale, we use the following logarithmic grid to resolve the low-frequency regime better
\begin{align}
\nu_k=\Delta \frac{2k-N_{\rm tot}}{N_{\rm tot}} \exp\Bigg\{\frac{|N_{\rm tot}-2k|-N_{\rm tot}}{S}\Bigg\},
\end{align}
with $k=0,1,\cdots N_{\rm tot}$. It distributes $N_{\rm tot}+1$ frequencies symmetrically around zero (Fermi-level) in interval $[-\Delta,\Delta]$. 
With parameter $S$, we can control the concentration of points around zero.
We choose the parameters such that the desired convergence ($10^{-10}$ in units of $\omega_0$) is achieved.  
Linear interpolation is used to evaluate the feedbacks $\epsilon^{\Lambda}(i\Lambda)$ and $\gamma^{\Lambda}(i\Lambda)$ 
on the right-hand side of Eqs.~(\ref{eq:flowepsilon}) and (\ref{eq:flowgamma}). 
The set of $2(N_{\rm tot}+1)$ differential equations can then be solved using standard adaptive routines.
\section{Phonon parameters}
\label{subsec:nb}
\begin{figure}[h]
 \centering
 \graphicspath{ {plots/}}
 \includegraphics[width=0.9\linewidth]{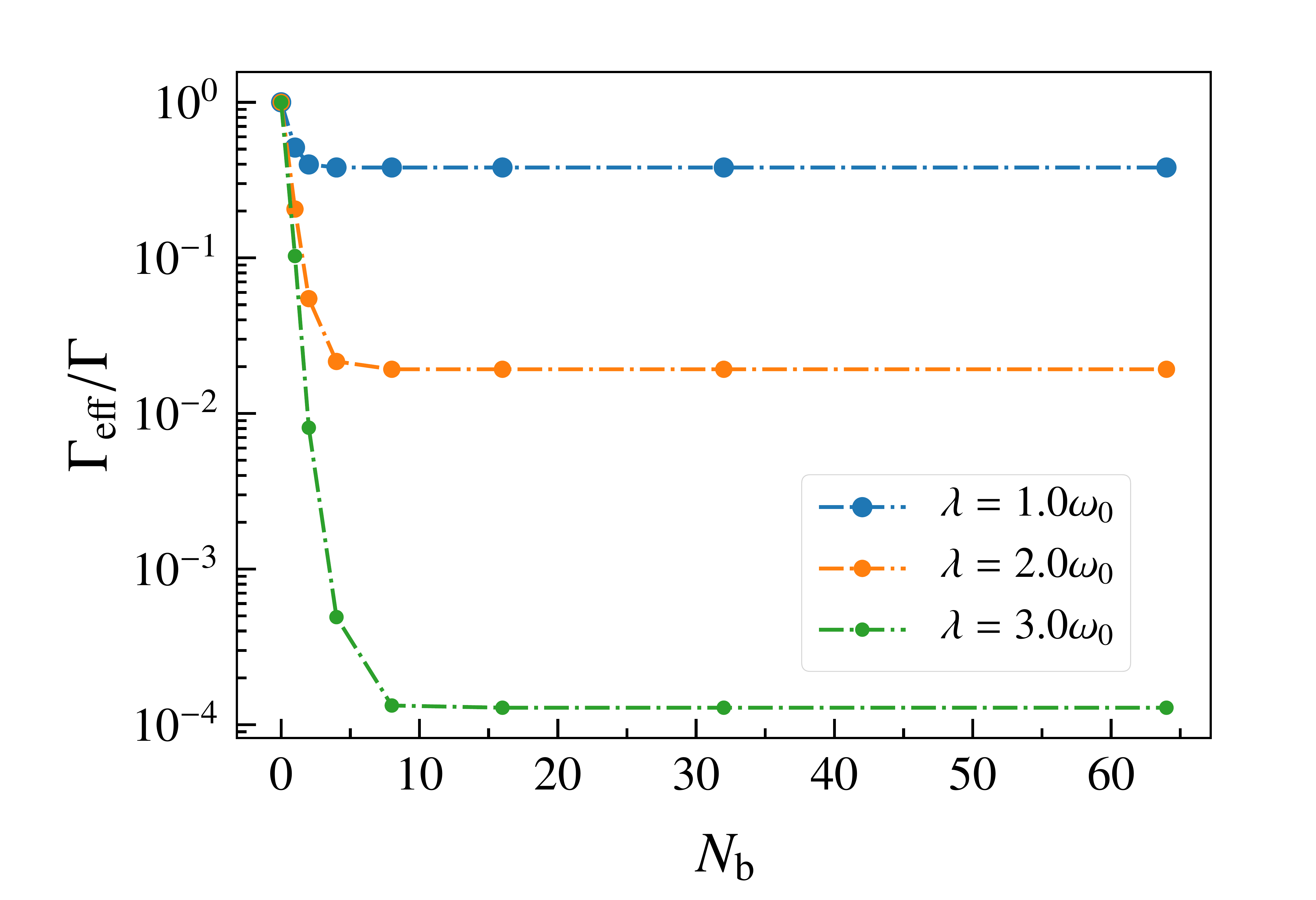}
 \caption{The ratio of the effective tunneling rate $\Gamma_{\rm eff}$ to the bare value $\Gamma$ as a function of $N_b$
 for different electron-phonon couplings in the antiadiabatic limit ($\omega_0=100\Gamma$).}
 \label{fig:nb}
\end{figure}

The minimum number of phonons required to resolve the low-energy behavior of the system depends on the strength of the electron-phonon coupling.
Figure \ref{fig:nb} shows the convergence of the effective tunneling
rate with the number of phonons retained. We used $N_b=40$ phonons
throughout, which suffices to obtain converged results for all parameters used.
\section{NRG and perturbation theory comparisons for the $T=0$ spectral function}
\label{subsec:spec-t0-comparisons}

\begin{figure}[!htbp]
 \centering
 \graphicspath{ {plots/}}
 \includegraphics[width=0.9\linewidth]{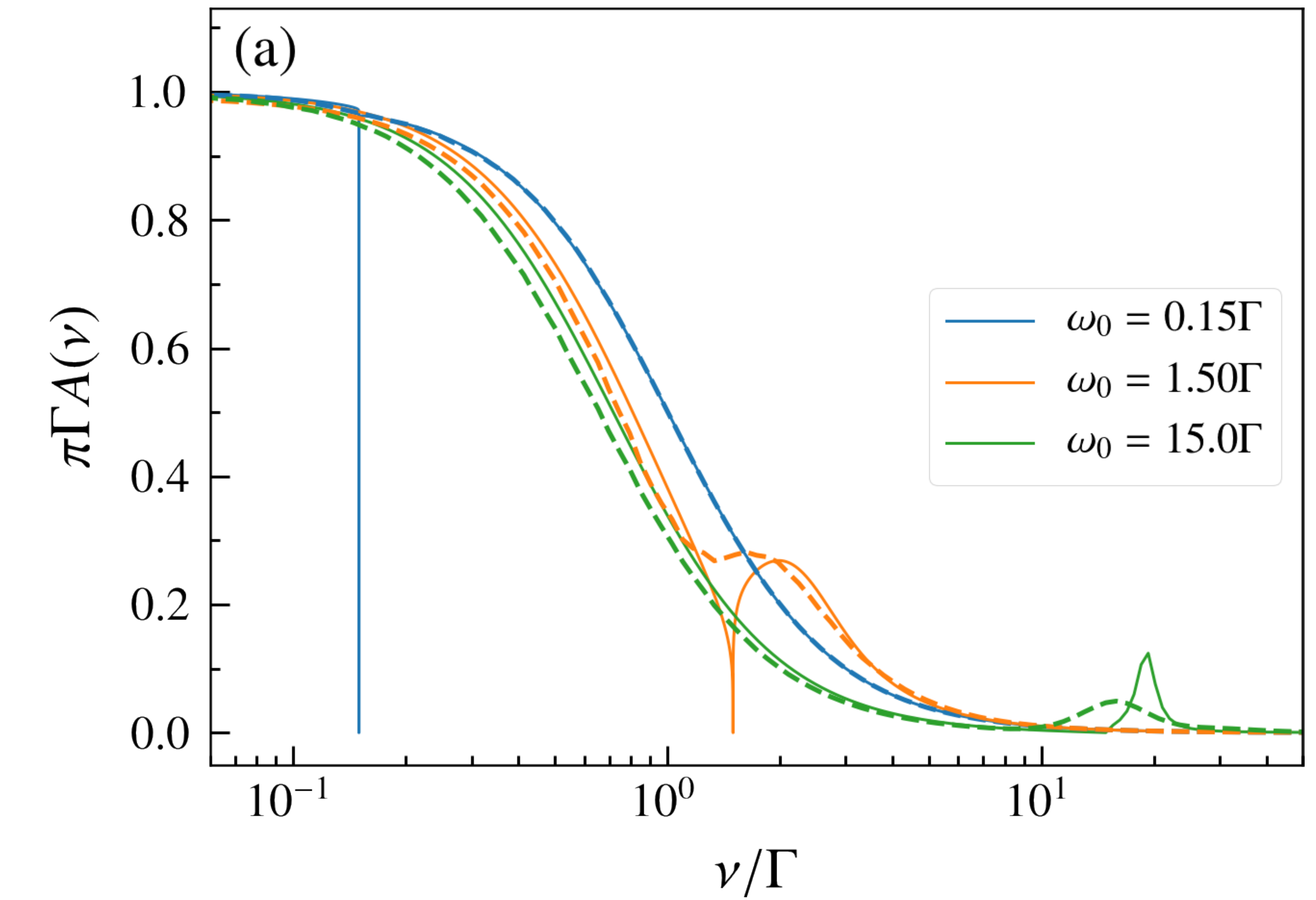}
\includegraphics[width=0.9\linewidth]{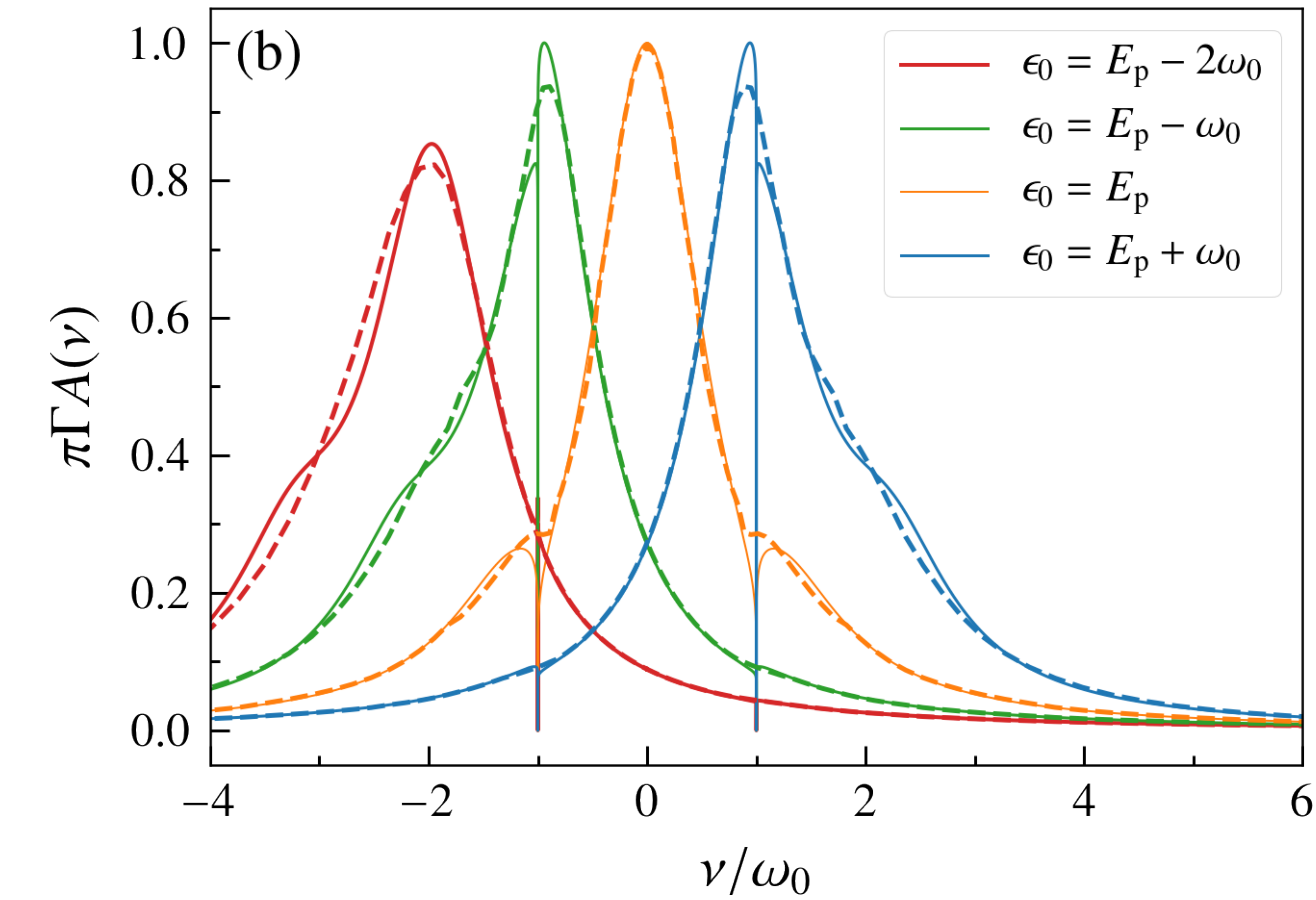}
 \caption{Comparison of the molecular 
spectral function obtained from perturbation theory (solid lines) and NRG (dashed lines).
(a) shows the spectral function at the particle-hole symmetric point 
$\epsilon_0=E_{\rm p}$ for different $\Gamma/\omega_0$ 
at a fixed electron-phonon coupling $\lambda=0.7\omega_0$ as a function of $\nu/\Gamma$. Note the logarithmic x-axis scale.
(b) depicts the spectral function at different gate voltages for 
$\omega_0=1.5\Gamma$ and $\lambda=0.5\omega_0$ as a function of 
$\nu/\omega_0$.} 
 \label{fig:spec}
\end{figure}

In the FRG approach, as formulated here, we have access to the propagator of the molecular
level only in Matsubara space and hence in order to obtain its spectral function,
we have to perform an analytic continuation to the real axis; this constitutes an ill-posed problem.
It becomes an obstacle as the self-energy at the end of the RG flow is
known only numerically. This problem was avoided in the calculation of
the conductance in Sec.~\ref{subsec:awayph} by using a continued fraction expansion of the
Fermi function. While we can
obtain results for spectral functions within FRG by performing
the analytic continuation numerically  via a P\'ade approximation,
we found this to be a quite unstable procedure in the sense that the results strongly depend on the number of data points and the frequency grid. 
An alternative, avoiding analytic continuation altogether, which we plan to follow in the future, is to use
FRG within the Keldysch formalism, which would also allow accessing
nonequilibrium.\cite{Jakobs07,Karrasch10,Laakso14} In order,
nevertheless, to compare our results for NRG spectral functions with
another method, we show here comparisons to lowest order perturbation theory. 

Figure~\ref{fig:spec} (a) shows the comparison of the spectral function 
computed within NRG and perturbation theory at particle-hole symmetry as 
we go from the adiabatic limit to the antiadiabatic one. We see that the central peak 
gets narrower, reflecting the suppression of tunneling processes. In addition, distinctive satellite peaks at multiples of 
$\omega_0$ start to form.
In the antiadiabatic limit, the width of the spectral function calculated from the two approaches do not quite match
in agreement to our previous discussion of the effective tunneling
coupling $\Gamma_{\rm eff}$ [see Fig.~\ref{fig:geffsep} (b)]. 

As discussed in connection with Eq.~(\ref{eq:fockreal}) the
perturbative retarded self-energy shows spurious logarithmic divergencies as $\nu= \pm \omega_0$. 
These lead to zeros of the spectral function which manifest as rather
sharp dips in Fig.~\ref{fig:spec}. These sharp features are
artifacts of the perturbation theory and are absent
in the NRG results. Similar artifacts of perturbation theory were found for the spinful 
Anderson-Holstein model by comparison to spectral functions obtained from Keldysh FRG (in equilibrium).\cite{Laakso14}  
For $\epsilon_0 > E_{\rm p}$ ($\epsilon_0 < E_{\rm p}$) the central peak, located at $\nu=0$ for particle-hole
symmetry, is shifted to the right (left). This effect is captured
rather accurately by perturbation theory for ${E_{\rm p}}/{\Gamma}<{\pi}/{2}$ [Fig.~\ref{fig:spec} (b)].

\bibliography{ref}

\begin{thebibliography}{47}%
\makeatletter
\providecommand \@ifxundefined [1]{%
 \@ifx{#1\undefined}
}%
\providecommand \@ifnum [1]{%
 \ifnum #1\expandafter \@firstoftwo
 \else \expandafter \@secondoftwo
 \fi
}%
\providecommand \@ifx [1]{%
 \ifx #1\expandafter \@firstoftwo
 \else \expandafter \@secondoftwo
 \fi
}%
\providecommand \natexlab [1]{#1}%
\providecommand \enquote  [1]{``#1''}%
\providecommand \bibnamefont  [1]{#1}%
\providecommand \bibfnamefont [1]{#1}%
\providecommand \citenamefont [1]{#1}%
\providecommand \href@noop [0]{\@secondoftwo}%
\providecommand \href [0]{\begingroup \@sanitize@url \@href}%
\providecommand \@href[1]{\@@startlink{#1}\@@href}%
\providecommand \@@href[1]{\endgroup#1\@@endlink}%
\providecommand \@sanitize@url [0]{\catcode `\\12\catcode `\$12\catcode
  `\&12\catcode `\#12\catcode `\^12\catcode `\_12\catcode `\%12\relax}%
\providecommand \@@startlink[1]{}%
\providecommand \@@endlink[0]{}%
\providecommand \url  [0]{\begingroup\@sanitize@url \@url }%
\providecommand \@url [1]{\endgroup\@href {#1}{\urlprefix }}%
\providecommand \urlprefix  [0]{URL }%
\providecommand \Eprint [0]{\href }%
\providecommand \doibase [0]{http://dx.doi.org/}%
\providecommand \selectlanguage [0]{\@gobble}%
\providecommand \bibinfo  [0]{\@secondoftwo}%
\providecommand \bibfield  [0]{\@secondoftwo}%
\providecommand \translation [1]{[#1]}%
\providecommand \BibitemOpen [0]{}%
\providecommand \bibitemStop [0]{}%
\providecommand \bibitemNoStop [0]{.\EOS\space}%
\providecommand \EOS [0]{\spacefactor3000\relax}%
\providecommand \BibitemShut  [1]{\csname bibitem#1\endcsname}%
\let\auto@bib@innerbib\@empty
\bibitem [{\citenamefont {Mahan}(2000)}]{Mahan90}%
  \BibitemOpen
  \bibfield  {author} {\bibinfo {author} {\bibfnamefont {G.}~\bibnamefont
  {Mahan}},\ }\href@noop {} {\emph {\bibinfo {title} {Many-Particle Physics}}}\
  (\bibinfo  {publisher} {Springer US},\ \bibinfo {address} {Boston, MA},\
  \bibinfo {year} {2000})\BibitemShut {NoStop}%
\bibitem [{\citenamefont {Mitra}\ \emph {et~al.}(2004)\citenamefont {Mitra},
  \citenamefont {Aleiner},\ and\ \citenamefont {Millis}}]{Mitra04}%
  \BibitemOpen
  \bibfield  {author} {\bibinfo {author} {\bibfnamefont {A.}~\bibnamefont
  {Mitra}}, \bibinfo {author} {\bibfnamefont {I.}~\bibnamefont {Aleiner}}, \
  and\ \bibinfo {author} {\bibfnamefont {A.~J.}\ \bibnamefont {Millis}},\
  }\href {\doibase 10.1103/PhysRevB.69.245302} {\bibfield  {journal} {\bibinfo
  {journal} {Phys. Rev. B}\ }\textbf {\bibinfo {volume} {69}},\ \bibinfo
  {pages} {245302} (\bibinfo {year} {2004})}\BibitemShut {NoStop}%
\bibitem [{\citenamefont {Sherrington}\ and\ \citenamefont {von
  Moln\`ar}(1975)}]{Sherrington75}%
  \BibitemOpen
  \bibfield  {author} {\bibinfo {author} {\bibfnamefont {D.}~\bibnamefont
  {Sherrington}}\ and\ \bibinfo {author} {\bibfnamefont {S.}~\bibnamefont {von
  Moln\`ar}},\ }\href {\doibase 10.1016/0038-1098(75)90843-1} {\bibfield
  {journal} {\bibinfo  {journal} {Solid State Communications}\ }\textbf
  {\bibinfo {volume} {16}},\ \bibinfo {pages} {1347 } (\bibinfo {year}
  {1975})}\BibitemShut {NoStop}%
\bibitem [{\citenamefont {Hewson}\ and\ \citenamefont
  {Newns}(1979)}]{Hewson79}%
  \BibitemOpen
  \bibfield  {author} {\bibinfo {author} {\bibfnamefont {A.~C.}\ \bibnamefont
  {Hewson}}\ and\ \bibinfo {author} {\bibfnamefont {D.~M.}\ \bibnamefont
  {Newns}},\ }\href {http://stacks.iop.org/0022-3719/12/i=9/a=009} {\bibfield
  {journal} {\bibinfo  {journal} {Journal of Physics C: Solid State Physics}\
  }\textbf {\bibinfo {volume} {12}},\ \bibinfo {pages} {1665} (\bibinfo {year}
  {1979})}\BibitemShut {NoStop}%
\bibitem [{\citenamefont {Hewson}\ and\ \citenamefont
  {Newns}(1980)}]{Hewson80}%
  \BibitemOpen
  \bibfield  {author} {\bibinfo {author} {\bibfnamefont {A.~C.}\ \bibnamefont
  {Hewson}}\ and\ \bibinfo {author} {\bibfnamefont {D.~M.}\ \bibnamefont
  {Newns}},\ }\href {http://stacks.iop.org/0022-3719/13/i=24/a=011} {\bibfield
  {journal} {\bibinfo  {journal} {Journal of Physics C: Solid State Physics}\
  }\textbf {\bibinfo {volume} {13}},\ \bibinfo {pages} {4477} (\bibinfo {year}
  {1980})}\BibitemShut {NoStop}%
\bibitem [{\citenamefont {Hewson}\ and\ \citenamefont
  {Meyer}(2002)}]{Hewson02}%
  \BibitemOpen
  \bibfield  {author} {\bibinfo {author} {\bibfnamefont {A.~C.}\ \bibnamefont
  {Hewson}}\ and\ \bibinfo {author} {\bibfnamefont {D.}~\bibnamefont {Meyer}},\
  }\href {http://stacks.iop.org/0953-8984/14/i=3/a=312} {\bibfield  {journal}
  {\bibinfo  {journal} {Journal of Physics: Condensed Matter}\ }\textbf
  {\bibinfo {volume} {14}},\ \bibinfo {pages} {427} (\bibinfo {year}
  {2002})}\BibitemShut {NoStop}%
\bibitem [{\citenamefont {Cornaglia}\ \emph {et~al.}(2004)\citenamefont
  {Cornaglia}, \citenamefont {Ness},\ and\ \citenamefont
  {Grempel}}]{Cornaglia04}%
  \BibitemOpen
  \bibfield  {author} {\bibinfo {author} {\bibfnamefont {P.~S.}\ \bibnamefont
  {Cornaglia}}, \bibinfo {author} {\bibfnamefont {H.}~\bibnamefont {Ness}}, \
  and\ \bibinfo {author} {\bibfnamefont {D.~R.}\ \bibnamefont {Grempel}},\
  }\href {\doibase 10.1103/PhysRevLett.93.147201} {\bibfield  {journal}
  {\bibinfo  {journal} {Phys. Rev. Lett.}\ }\textbf {\bibinfo {volume} {93}},\
  \bibinfo {pages} {147201} (\bibinfo {year} {2004})}\BibitemShut {NoStop}%
\bibitem [{\citenamefont {Paaske}\ and\ \citenamefont
  {Flensberg}(2005)}]{Paaske05}%
  \BibitemOpen
  \bibfield  {author} {\bibinfo {author} {\bibfnamefont {J.}~\bibnamefont
  {Paaske}}\ and\ \bibinfo {author} {\bibfnamefont {K.}~\bibnamefont
  {Flensberg}},\ }\href {\doibase 10.1103/PhysRevLett.94.176801} {\bibfield
  {journal} {\bibinfo  {journal} {Phys. Rev. Lett.}\ }\textbf {\bibinfo
  {volume} {94}},\ \bibinfo {pages} {176801} (\bibinfo {year}
  {2005})}\BibitemShut {NoStop}%
\bibitem [{\citenamefont {Galperin}\ \emph {et~al.}(2007)\citenamefont
  {Galperin}, \citenamefont {Ratner},\ and\ \citenamefont
  {Nitzan}}]{Galperin07}%
  \BibitemOpen
  \bibfield  {author} {\bibinfo {author} {\bibfnamefont {M.}~\bibnamefont
  {Galperin}}, \bibinfo {author} {\bibfnamefont {M.~A.}\ \bibnamefont
  {Ratner}}, \ and\ \bibinfo {author} {\bibfnamefont {A.}~\bibnamefont
  {Nitzan}},\ }\href {http://stacks.iop.org/0953-8984/19/i=10/a=103201}
  {\bibfield  {journal} {\bibinfo  {journal} {Journal of Physics: Condensed
  Matter}\ }\textbf {\bibinfo {volume} {19}},\ \bibinfo {pages} {103201}
  (\bibinfo {year} {2007})}\BibitemShut {NoStop}%
\bibitem [{\citenamefont {Hyldgaard}\ \emph {et~al.}(1994)\citenamefont
  {Hyldgaard}, \citenamefont {Hershfield}, \citenamefont {Davies},\ and\
  \citenamefont {Wilkins}}]{Hyldgaard94}%
  \BibitemOpen
  \bibfield  {author} {\bibinfo {author} {\bibfnamefont {P.}~\bibnamefont
  {Hyldgaard}}, \bibinfo {author} {\bibfnamefont {S.}~\bibnamefont
  {Hershfield}}, \bibinfo {author} {\bibfnamefont {J.}~\bibnamefont {Davies}},
  \ and\ \bibinfo {author} {\bibfnamefont {J.}~\bibnamefont {Wilkins}},\ }\href
  {\doibase 10.1006/aphy.1994.1106} {\bibfield  {journal} {\bibinfo  {journal}
  {Annals of Physics}\ }\textbf {\bibinfo {volume} {236}},\ \bibinfo {pages} {1
  } (\bibinfo {year} {1994})}\BibitemShut {NoStop}%
\bibitem [{\citenamefont {K\"onig}\ \emph {et~al.}(1996)\citenamefont
  {K\"onig}, \citenamefont {Schmid}, \citenamefont {Schoeller},\ and\
  \citenamefont {Sch\"on}}]{Koenig96}%
  \BibitemOpen
  \bibfield  {author} {\bibinfo {author} {\bibfnamefont {J.}~\bibnamefont
  {K\"onig}}, \bibinfo {author} {\bibfnamefont {J.}~\bibnamefont {Schmid}},
  \bibinfo {author} {\bibfnamefont {H.}~\bibnamefont {Schoeller}}, \ and\
  \bibinfo {author} {\bibfnamefont {G.}~\bibnamefont {Sch\"on}},\ }\href
  {\doibase 10.1103/PhysRevB.54.16820} {\bibfield  {journal} {\bibinfo
  {journal} {Phys. Rev. B}\ }\textbf {\bibinfo {volume} {54}},\ \bibinfo
  {pages} {16820} (\bibinfo {year} {1996})}\BibitemShut {NoStop}%
\bibitem [{\citenamefont {Braig}\ and\ \citenamefont
  {Flensberg}(2003)}]{Braig03}%
  \BibitemOpen
  \bibfield  {author} {\bibinfo {author} {\bibfnamefont {S.}~\bibnamefont
  {Braig}}\ and\ \bibinfo {author} {\bibfnamefont {K.}~\bibnamefont
  {Flensberg}},\ }\href {\doibase 10.1103/PhysRevB.68.205324} {\bibfield
  {journal} {\bibinfo  {journal} {Phys. Rev. B}\ }\textbf {\bibinfo {volume}
  {68}},\ \bibinfo {pages} {205324} (\bibinfo {year} {2003})}\BibitemShut
  {NoStop}%
\bibitem [{\citenamefont {Koch}\ and\ \citenamefont {von
  Oppen}(2005)}]{Koch05}%
  \BibitemOpen
  \bibfield  {author} {\bibinfo {author} {\bibfnamefont {J.}~\bibnamefont
  {Koch}}\ and\ \bibinfo {author} {\bibfnamefont {F.}~\bibnamefont {von
  Oppen}},\ }\href {\doibase 10.1103/PhysRevLett.94.206804} {\bibfield
  {journal} {\bibinfo  {journal} {Phys. Rev. Lett.}\ }\textbf {\bibinfo
  {volume} {94}},\ \bibinfo {pages} {206804} (\bibinfo {year}
  {2005})}\BibitemShut {NoStop}%
\bibitem [{\citenamefont {Han}(2006)}]{Han06}%
  \BibitemOpen
  \bibfield  {author} {\bibinfo {author} {\bibfnamefont {J.~E.}\ \bibnamefont
  {Han}},\ }\href {\doibase 10.1103/PhysRevB.73.125319} {\bibfield  {journal}
  {\bibinfo  {journal} {Phys. Rev. B}\ }\textbf {\bibinfo {volume} {73}},\
  \bibinfo {pages} {125319} (\bibinfo {year} {2006})}\BibitemShut {NoStop}%
\bibitem [{\citenamefont {Han}(2010)}]{Han10}%
  \BibitemOpen
  \bibfield  {author} {\bibinfo {author} {\bibfnamefont {J.~E.}\ \bibnamefont
  {Han}},\ }\href {\doibase 10.1103/PhysRevB.81.113106} {\bibfield  {journal}
  {\bibinfo  {journal} {Phys. Rev. B}\ }\textbf {\bibinfo {volume} {81}},\
  \bibinfo {pages} {113106} (\bibinfo {year} {2010})}\BibitemShut {NoStop}%
\bibitem [{\citenamefont {M\"uhlbacher}\ and\ \citenamefont
  {Rabani}(2008)}]{Muehlbacher08}%
  \BibitemOpen
  \bibfield  {author} {\bibinfo {author} {\bibfnamefont {L.}~\bibnamefont
  {M\"uhlbacher}}\ and\ \bibinfo {author} {\bibfnamefont {E.}~\bibnamefont
  {Rabani}},\ }\href {\doibase 10.1103/PhysRevLett.100.176403} {\bibfield
  {journal} {\bibinfo  {journal} {Phys. Rev. Lett.}\ }\textbf {\bibinfo
  {volume} {100}},\ \bibinfo {pages} {176403} (\bibinfo {year}
  {2008})}\BibitemShut {NoStop}%
\bibitem [{\citenamefont {Schir\'o}\ and\ \citenamefont
  {Fabrizio}(2009)}]{Schiro09}%
  \BibitemOpen
  \bibfield  {author} {\bibinfo {author} {\bibfnamefont {M.}~\bibnamefont
  {Schir\'o}}\ and\ \bibinfo {author} {\bibfnamefont {M.}~\bibnamefont
  {Fabrizio}},\ }\href {\doibase 10.1103/PhysRevB.79.153302} {\bibfield
  {journal} {\bibinfo  {journal} {Phys. Rev. B}\ }\textbf {\bibinfo {volume}
  {79}},\ \bibinfo {pages} {153302} (\bibinfo {year} {2009})}\BibitemShut
  {NoStop}%
\bibitem [{\citenamefont {Koch}\ \emph {et~al.}(2011)\citenamefont {Koch},
  \citenamefont {Loos}, \citenamefont {Alvermann},\ and\ \citenamefont
  {Fehske}}]{Koch11}%
  \BibitemOpen
  \bibfield  {author} {\bibinfo {author} {\bibfnamefont {T.}~\bibnamefont
  {Koch}}, \bibinfo {author} {\bibfnamefont {J.}~\bibnamefont {Loos}}, \bibinfo
  {author} {\bibfnamefont {A.}~\bibnamefont {Alvermann}}, \ and\ \bibinfo
  {author} {\bibfnamefont {H.}~\bibnamefont {Fehske}},\ }\href {\doibase
  10.1103/PhysRevB.84.125131} {\bibfield  {journal} {\bibinfo  {journal} {Phys.
  Rev. B}\ }\textbf {\bibinfo {volume} {84}},\ \bibinfo {pages} {125131}
  (\bibinfo {year} {2011})}\BibitemShut {NoStop}%
\bibitem [{\citenamefont {H\"utzen}\ \emph {et~al.}(2012)\citenamefont
  {H\"utzen}, \citenamefont {Weiss}, \citenamefont {Thorwart},\ and\
  \citenamefont {Egger}}]{Huetzen12}%
  \BibitemOpen
  \bibfield  {author} {\bibinfo {author} {\bibfnamefont {R.}~\bibnamefont
  {H\"utzen}}, \bibinfo {author} {\bibfnamefont {S.}~\bibnamefont {Weiss}},
  \bibinfo {author} {\bibfnamefont {M.}~\bibnamefont {Thorwart}}, \ and\
  \bibinfo {author} {\bibfnamefont {R.}~\bibnamefont {Egger}},\ }\href
  {\doibase 10.1103/PhysRevB.85.121408} {\bibfield  {journal} {\bibinfo
  {journal} {Phys. Rev. B}\ }\textbf {\bibinfo {volume} {85}},\ \bibinfo
  {pages} {121408} (\bibinfo {year} {2012})}\BibitemShut {NoStop}%
\bibitem [{\citenamefont {Jovchev}\ and\ \citenamefont
  {Anders}(2013)}]{Jovchev13}%
  \BibitemOpen
  \bibfield  {author} {\bibinfo {author} {\bibfnamefont {A.}~\bibnamefont
  {Jovchev}}\ and\ \bibinfo {author} {\bibfnamefont {F.~B.}\ \bibnamefont
  {Anders}},\ }\href {\doibase 10.1103/PhysRevB.87.195112} {\bibfield
  {journal} {\bibinfo  {journal} {Phys. Rev. B}\ }\textbf {\bibinfo {volume}
  {87}},\ \bibinfo {pages} {195112} (\bibinfo {year} {2013})}\BibitemShut
  {NoStop}%
\bibitem [{\citenamefont {Roura-Bas}\ \emph {et~al.}(2013)\citenamefont
  {Roura-Bas}, \citenamefont {Tosi},\ and\ \citenamefont
  {Aligia}}]{RouraBas13}%
  \BibitemOpen
  \bibfield  {author} {\bibinfo {author} {\bibfnamefont {P.}~\bibnamefont
  {Roura-Bas}}, \bibinfo {author} {\bibfnamefont {L.}~\bibnamefont {Tosi}}, \
  and\ \bibinfo {author} {\bibfnamefont {A.~A.}\ \bibnamefont {Aligia}},\
  }\href {\doibase 10.1103/PhysRevB.87.195136} {\bibfield  {journal} {\bibinfo
  {journal} {Phys. Rev. B}\ }\textbf {\bibinfo {volume} {87}},\ \bibinfo
  {pages} {195136} (\bibinfo {year} {2013})}\BibitemShut {NoStop}%
\bibitem [{\citenamefont {Laakso}\ \emph {et~al.}(2014)\citenamefont {Laakso},
  \citenamefont {Kennes}, \citenamefont {Jakobs},\ and\ \citenamefont
  {Meden}}]{Laakso14}%
  \BibitemOpen
  \bibfield  {author} {\bibinfo {author} {\bibfnamefont {M.~A.}\ \bibnamefont
  {Laakso}}, \bibinfo {author} {\bibfnamefont {D.~M.}\ \bibnamefont {Kennes}},
  \bibinfo {author} {\bibfnamefont {S.~G.}\ \bibnamefont {Jakobs}}, \ and\
  \bibinfo {author} {\bibfnamefont {V.}~\bibnamefont {Meden}},\ }\href
  {http://stacks.iop.org/1367-2630/16/i=2/a=023007} {\bibfield  {journal}
  {\bibinfo  {journal} {New Journal of Physics}\ }\textbf {\bibinfo {volume}
  {16}},\ \bibinfo {pages} {023007} (\bibinfo {year} {2014})}\BibitemShut
  {NoStop}%
\bibitem [{\citenamefont {Albrecht}\ \emph {et~al.}(2013)\citenamefont
  {Albrecht}, \citenamefont {Martin-Rodero}, \citenamefont {Monreal},
  \citenamefont {M\"uhlbacher},\ and\ \citenamefont
  {Levy~Yeyati}}]{Albrecht13}%
  \BibitemOpen
  \bibfield  {author} {\bibinfo {author} {\bibfnamefont {K.~F.}\ \bibnamefont
  {Albrecht}}, \bibinfo {author} {\bibfnamefont {A.}~\bibnamefont
  {Martin-Rodero}}, \bibinfo {author} {\bibfnamefont {R.~C.}\ \bibnamefont
  {Monreal}}, \bibinfo {author} {\bibfnamefont {L.}~\bibnamefont
  {M\"uhlbacher}}, \ and\ \bibinfo {author} {\bibfnamefont {A.}~\bibnamefont
  {Levy~Yeyati}},\ }\href {\doibase 10.1103/PhysRevB.87.085127} {\bibfield
  {journal} {\bibinfo  {journal} {Phys. Rev. B}\ }\textbf {\bibinfo {volume}
  {87}},\ \bibinfo {pages} {085127} (\bibinfo {year} {2013})}\BibitemShut
  {NoStop}%
\bibitem [{\citenamefont {Leturcq}\ \emph {et~al.}(2009)\citenamefont
  {Leturcq}, \citenamefont {Stampfer}, \citenamefont {Inderbitzin},
  \citenamefont {Durrer}, \citenamefont {Hierold}, \citenamefont {Mariani},
  \citenamefont {Schultz}, \citenamefont {von Oppen},\ and\ \citenamefont
  {Ensslin}}]{Leturcq09}%
  \BibitemOpen
  \bibfield  {author} {\bibinfo {author} {\bibfnamefont {R.}~\bibnamefont
  {Leturcq}}, \bibinfo {author} {\bibfnamefont {C.}~\bibnamefont {Stampfer}},
  \bibinfo {author} {\bibfnamefont {K.}~\bibnamefont {Inderbitzin}}, \bibinfo
  {author} {\bibfnamefont {L.}~\bibnamefont {Durrer}}, \bibinfo {author}
  {\bibfnamefont {C.}~\bibnamefont {Hierold}}, \bibinfo {author} {\bibfnamefont
  {E.}~\bibnamefont {Mariani}}, \bibinfo {author} {\bibfnamefont {M.~G.}\
  \bibnamefont {Schultz}}, \bibinfo {author} {\bibfnamefont {F.}~\bibnamefont
  {von Oppen}}, \ and\ \bibinfo {author} {\bibfnamefont {K.}~\bibnamefont
  {Ensslin}},\ }\href {\doibase 10.1038/nphys1234} {\bibfield  {journal}
  {\bibinfo  {journal} {Nat Phys}\ }\textbf {\bibinfo {volume} {5}},\ \bibinfo
  {pages} {327} (\bibinfo {year} {2009})}\BibitemShut {NoStop}%
\bibitem [{\citenamefont {Lang}\ and\ \citenamefont {Firsov}(1962)}]{Lang62}%
  \BibitemOpen
  \bibfield  {author} {\bibinfo {author} {\bibfnamefont {I.}~\bibnamefont
  {Lang}}\ and\ \bibinfo {author} {\bibfnamefont {Y.}~\bibnamefont {Firsov}},\
  }\href@noop {} {\bibfield  {journal} {\bibinfo  {journal} {Zh. Eksp. Teor.
  Fiz.}\ }\textbf {\bibinfo {volume} {43}},\ \bibinfo {pages} {1843} (\bibinfo
  {year} {1962})},\ \bibinfo {note}
  {[\href{http://www.jetp.ac.ru/cgi-bin/dn/e_016_05_1301.pdf}{Sov. Phys.-JETP
  {\bf 16}, 1301 (1963)}]}\BibitemShut {NoStop}%
\bibitem [{\citenamefont {Eidelstein}\ \emph {et~al.}(2013)\citenamefont
  {Eidelstein}, \citenamefont {Goberman},\ and\ \citenamefont
  {Schiller}}]{Eidelstein13}%
  \BibitemOpen
  \bibfield  {author} {\bibinfo {author} {\bibfnamefont {E.}~\bibnamefont
  {Eidelstein}}, \bibinfo {author} {\bibfnamefont {D.}~\bibnamefont
  {Goberman}}, \ and\ \bibinfo {author} {\bibfnamefont {A.}~\bibnamefont
  {Schiller}},\ }\href {\doibase 10.1103/PhysRevB.87.075319} {\bibfield
  {journal} {\bibinfo  {journal} {Phys. Rev. B}\ }\textbf {\bibinfo {volume}
  {87}},\ \bibinfo {pages} {075319} (\bibinfo {year} {2013})}\BibitemShut
  {NoStop}%
\bibitem [{\citenamefont {Schlottmann}(1980)}]{Schlottmann80}%
  \BibitemOpen
  \bibfield  {author} {\bibinfo {author} {\bibfnamefont {P.}~\bibnamefont
  {Schlottmann}},\ }\href {\doibase 10.1103/PhysRevB.22.613} {\bibfield
  {journal} {\bibinfo  {journal} {Phys. Rev. B}\ }\textbf {\bibinfo {volume}
  {22}},\ \bibinfo {pages} {613} (\bibinfo {year} {1980})}\BibitemShut
  {NoStop}%
\bibitem [{\citenamefont {Schlottmann}(1982)}]{Schlottmann82}%
  \BibitemOpen
  \bibfield  {author} {\bibinfo {author} {\bibfnamefont {P.}~\bibnamefont
  {Schlottmann}},\ }\href {\doibase 10.1103/PhysRevB.25.4815} {\bibfield
  {journal} {\bibinfo  {journal} {Phys. Rev. B}\ }\textbf {\bibinfo {volume}
  {25}},\ \bibinfo {pages} {4815} (\bibinfo {year} {1982})}\BibitemShut
  {NoStop}%
\bibitem [{\citenamefont {Khedri}\ \emph {et~al.}(2017)\citenamefont {Khedri},
  \citenamefont {Meden},\ and\ \citenamefont {Costi}}]{Khedri17b}%
  \BibitemOpen
  \bibfield  {author} {\bibinfo {author} {\bibfnamefont {A.}~\bibnamefont
  {Khedri}}, \bibinfo {author} {\bibfnamefont {V.}~\bibnamefont {Meden}}, \
  and\ \bibinfo {author} {\bibfnamefont {T.~A.}\ \bibnamefont {Costi}},\ }\href
  {\doibase 10.1103/PhysRevB.96.195156} {\bibfield  {journal} {\bibinfo
  {journal} {Phys. Rev. B}\ }\textbf {\bibinfo {volume} {96}},\ \bibinfo
  {pages} {195156} (\bibinfo {year} {2017})}\BibitemShut {NoStop}%
\bibitem [{\citenamefont {Metzner}\ \emph {et~al.}(2012)\citenamefont
  {Metzner}, \citenamefont {Salmhofer}, \citenamefont {Honerkamp},
  \citenamefont {Meden},\ and\ \citenamefont {Sch\"onhammer}}]{Metzner12}%
  \BibitemOpen
  \bibfield  {author} {\bibinfo {author} {\bibfnamefont {W.}~\bibnamefont
  {Metzner}}, \bibinfo {author} {\bibfnamefont {M.}~\bibnamefont {Salmhofer}},
  \bibinfo {author} {\bibfnamefont {C.}~\bibnamefont {Honerkamp}}, \bibinfo
  {author} {\bibfnamefont {V.}~\bibnamefont {Meden}}, \ and\ \bibinfo {author}
  {\bibfnamefont {K.}~\bibnamefont {Sch\"onhammer}},\ }\href {\doibase
  10.1103/RevModPhys.84.299} {\bibfield  {journal} {\bibinfo  {journal} {Rev.
  Mod. Phys.}\ }\textbf {\bibinfo {volume} {84}},\ \bibinfo {pages} {299}
  (\bibinfo {year} {2012})}\BibitemShut {NoStop}%
\bibitem [{\citenamefont {Karrasch}\ \emph
  {et~al.}(2010{\natexlab{a}})\citenamefont {Karrasch}, \citenamefont
  {Pletyukhov}, \citenamefont {Borda},\ and\ \citenamefont
  {Meden}}]{Karrasch10}%
  \BibitemOpen
  \bibfield  {author} {\bibinfo {author} {\bibfnamefont {C.}~\bibnamefont
  {Karrasch}}, \bibinfo {author} {\bibfnamefont {M.}~\bibnamefont
  {Pletyukhov}}, \bibinfo {author} {\bibfnamefont {L.}~\bibnamefont {Borda}}, \
  and\ \bibinfo {author} {\bibfnamefont {V.}~\bibnamefont {Meden}},\ }\href
  {\doibase 10.1103/PhysRevB.81.125122} {\bibfield  {journal} {\bibinfo
  {journal} {Phys. Rev. B}\ }\textbf {\bibinfo {volume} {81}},\ \bibinfo
  {pages} {125122} (\bibinfo {year} {2010}{\natexlab{a}})}\BibitemShut
  {NoStop}%
\bibitem [{\citenamefont {Karrasch}\ \emph {et~al.}(2006)\citenamefont
  {Karrasch}, \citenamefont {Enss},\ and\ \citenamefont {Meden}}]{Karrasch06}%
  \BibitemOpen
  \bibfield  {author} {\bibinfo {author} {\bibfnamefont {C.}~\bibnamefont
  {Karrasch}}, \bibinfo {author} {\bibfnamefont {T.}~\bibnamefont {Enss}}, \
  and\ \bibinfo {author} {\bibfnamefont {V.}~\bibnamefont {Meden}},\ }\href
  {\doibase 10.1103/PhysRevB.73.235337} {\bibfield  {journal} {\bibinfo
  {journal} {Phys. Rev. B}\ }\textbf {\bibinfo {volume} {73}},\ \bibinfo
  {pages} {235337} (\bibinfo {year} {2006})}\BibitemShut {NoStop}%
\bibitem [{\citenamefont {Kennes}(2014)}]{Kennes14}%
  \BibitemOpen
  \bibfield  {author} {\bibinfo {author} {\bibfnamefont {D.~M.}\ \bibnamefont
  {Kennes}},\ }\href {https://publications.rwth-aachen.de/record/459464} {Ph.D.
  thesis},\ \bibinfo  {school} {RWTH Aachen University} (\bibinfo {year}
  {2014})\BibitemShut {NoStop}%
\bibitem [{\citenamefont {Campo}\ and\ \citenamefont
  {Oliveira}(2005)}]{CampoOliveira05}%
  \BibitemOpen
  \bibfield  {author} {\bibinfo {author} {\bibfnamefont {V.~L.}\ \bibnamefont
  {Campo}}\ and\ \bibinfo {author} {\bibfnamefont {L.~N.}\ \bibnamefont
  {Oliveira}},\ }\href {\doibase 10.1103/PhysRevB.72.104432} {\bibfield
  {journal} {\bibinfo  {journal} {Phys. Rev. B}\ }\textbf {\bibinfo {volume}
  {72}},\ \bibinfo {pages} {104432} (\bibinfo {year} {2005})}\BibitemShut
  {NoStop}%
\bibitem [{\citenamefont {Wilson}(1975)}]{Wilson75}%
  \BibitemOpen
  \bibfield  {author} {\bibinfo {author} {\bibfnamefont {K.~G.}\ \bibnamefont
  {Wilson}},\ }\href {\doibase 10.1103/RevModPhys.47.773} {\bibfield  {journal}
  {\bibinfo  {journal} {Rev. Mod. Phys.}\ }\textbf {\bibinfo {volume} {47}},\
  \bibinfo {pages} {773} (\bibinfo {year} {1975})}\BibitemShut {NoStop}%
\bibitem [{\citenamefont {Krishna-murthy}\ \emph {et~al.}(1980)\citenamefont
  {Krishna-murthy}, \citenamefont {Wilkins},\ and\ \citenamefont
  {Wilson}}]{Krishnamurthy80}%
  \BibitemOpen
  \bibfield  {author} {\bibinfo {author} {\bibfnamefont {H.~R.}\ \bibnamefont
  {Krishna-murthy}}, \bibinfo {author} {\bibfnamefont {J.~W.}\ \bibnamefont
  {Wilkins}}, \ and\ \bibinfo {author} {\bibfnamefont {K.~G.}\ \bibnamefont
  {Wilson}},\ }\href {\doibase 10.1103/PhysRevB.21.1003} {\bibfield  {journal}
  {\bibinfo  {journal} {Phys. Rev. B}\ }\textbf {\bibinfo {volume} {21}},\
  \bibinfo {pages} {1003} (\bibinfo {year} {1980})}\BibitemShut {NoStop}%
\bibitem [{\citenamefont {Bulla}\ \emph {et~al.}(2008)\citenamefont {Bulla},
  \citenamefont {Costi},\ and\ \citenamefont {Pruschke}}]{Bulla08}%
  \BibitemOpen
  \bibfield  {author} {\bibinfo {author} {\bibfnamefont {R.}~\bibnamefont
  {Bulla}}, \bibinfo {author} {\bibfnamefont {T.~A.}\ \bibnamefont {Costi}}, \
  and\ \bibinfo {author} {\bibfnamefont {T.}~\bibnamefont {Pruschke}},\ }\href
  {\doibase 10.1103/RevModPhys.80.395} {\bibfield  {journal} {\bibinfo
  {journal} {Rev. Mod. Phys.}\ }\textbf {\bibinfo {volume} {80}},\ \bibinfo
  {pages} {395} (\bibinfo {year} {2008})}\BibitemShut {NoStop}%
\bibitem [{\citenamefont {Ingersent}(2015)}]{Ingersent15}%
  \BibitemOpen
  \bibfield  {author} {\bibinfo {author} {\bibfnamefont {K.}~\bibnamefont
  {Ingersent}},\ }in\ \href {http://juser.fz-juelich.de/record/205123} {\emph
  {\bibinfo {booktitle} {{M}any-{B}ody {P}hysics: {F}rom {K}ondo to
  {H}ubbard}}},\ \bibinfo {series} {Modeling and Simulation}, Vol.~\bibinfo
  {volume} {5},\ \bibinfo {editor} {edited by\ \bibinfo {editor} {\bibfnamefont
  {E.}~\bibnamefont {Pavarini}}, \bibinfo {editor} {\bibfnamefont
  {E.}~\bibnamefont {Koch}}, \ and\ \bibinfo {editor} {\bibfnamefont
  {P.}~\bibnamefont {Coleman}}}\ (\bibinfo  {publisher} {Forschungszentrum
  J{\"u}lich GmbH Zentralbibliothek, Verlag},\ \bibinfo {address}
  {J{\"u}lich},\ \bibinfo {year} {2015})\ Chap.~\bibinfo {chapter}
  {6}\BibitemShut {NoStop}%
\bibitem [{\citenamefont {Bulla}\ \emph {et~al.}(1998)\citenamefont {Bulla},
  \citenamefont {Hewson},\ and\ \citenamefont {Pruschke}}]{Bulla98}%
  \BibitemOpen
  \bibfield  {author} {\bibinfo {author} {\bibfnamefont {R.}~\bibnamefont
  {Bulla}}, \bibinfo {author} {\bibfnamefont {A.~C.}\ \bibnamefont {Hewson}}, \
  and\ \bibinfo {author} {\bibfnamefont {T.}~\bibnamefont {Pruschke}},\ }\href
  {http://stacks.iop.org/0953-8984/10/i=37/a=021} {\bibfield  {journal}
  {\bibinfo  {journal} {Journal of Physics: Condensed Matter}\ }\textbf
  {\bibinfo {volume} {10}},\ \bibinfo {pages} {8365} (\bibinfo {year}
  {1998})}\BibitemShut {NoStop}%
\bibitem [{\citenamefont {Vinkler}\ \emph {et~al.}(2012)\citenamefont
  {Vinkler}, \citenamefont {Schiller},\ and\ \citenamefont
  {Andrei}}]{Vinkler12}%
  \BibitemOpen
  \bibfield  {author} {\bibinfo {author} {\bibfnamefont {Y.}~\bibnamefont
  {Vinkler}}, \bibinfo {author} {\bibfnamefont {A.}~\bibnamefont {Schiller}}, \
  and\ \bibinfo {author} {\bibfnamefont {N.}~\bibnamefont {Andrei}},\ }\href
  {\doibase 10.1103/PhysRevB.85.035411} {\bibfield  {journal} {\bibinfo
  {journal} {Phys. Rev. B}\ }\textbf {\bibinfo {volume} {85}},\ \bibinfo
  {pages} {035411} (\bibinfo {year} {2012})}\BibitemShut {NoStop}%
\bibitem [{\citenamefont {Meir}\ and\ \citenamefont {Wingreen}(1992)}]{Meir92}%
  \BibitemOpen
  \bibfield  {author} {\bibinfo {author} {\bibfnamefont {Y.}~\bibnamefont
  {Meir}}\ and\ \bibinfo {author} {\bibfnamefont {N.~S.}\ \bibnamefont
  {Wingreen}},\ }\href {\doibase 10.1103/PhysRevLett.68.2512} {\bibfield
  {journal} {\bibinfo  {journal} {Phys. Rev. Lett.}\ }\textbf {\bibinfo
  {volume} {68}},\ \bibinfo {pages} {2512} (\bibinfo {year}
  {1992})}\BibitemShut {NoStop}%
\bibitem [{\citenamefont {Ozaki}(2007)}]{Ozaki07}%
  \BibitemOpen
  \bibfield  {author} {\bibinfo {author} {\bibfnamefont {T.}~\bibnamefont
  {Ozaki}},\ }\href {\doibase 10.1103/PhysRevB.75.035123} {\bibfield  {journal}
  {\bibinfo  {journal} {Phys. Rev. B}\ }\textbf {\bibinfo {volume} {75}},\
  \bibinfo {pages} {035123} (\bibinfo {year} {2007})}\BibitemShut {NoStop}%
\bibitem [{\citenamefont {Karrasch}\ \emph
  {et~al.}(2010{\natexlab{b}})\citenamefont {Karrasch}, \citenamefont {Meden},\
  and\ \citenamefont {Sch\"onhammer}}]{KarraschOzaki10}%
  \BibitemOpen
  \bibfield  {author} {\bibinfo {author} {\bibfnamefont {C.}~\bibnamefont
  {Karrasch}}, \bibinfo {author} {\bibfnamefont {V.}~\bibnamefont {Meden}}, \
  and\ \bibinfo {author} {\bibfnamefont {K.}~\bibnamefont {Sch\"onhammer}},\
  }\href {\doibase 10.1103/PhysRevB.82.125114} {\bibfield  {journal} {\bibinfo
  {journal} {Phys. Rev. B}\ }\textbf {\bibinfo {volume} {82}},\ \bibinfo
  {pages} {125114} (\bibinfo {year} {2010}{\natexlab{b}})}\BibitemShut
  {NoStop}%
\bibitem [{\citenamefont {Mahan}\ \emph {et~al.}(1997)\citenamefont {Mahan},
  \citenamefont {Sales},\ and\ \citenamefont {Sharp}}]{Mahan97}%
  \BibitemOpen
  \bibfield  {author} {\bibinfo {author} {\bibfnamefont {G.}~\bibnamefont
  {Mahan}}, \bibinfo {author} {\bibfnamefont {B.}~\bibnamefont {Sales}}, \ and\
  \bibinfo {author} {\bibfnamefont {J.}~\bibnamefont {Sharp}},\ }\href
  {http://dx.doi.org/10.1063/1.881752} {\bibfield  {journal} {\bibinfo
  {journal} {Phys. Today}\ }\textbf {\bibinfo {volume} {50}},\ \bibinfo {pages}
  {42} (\bibinfo {year} {1997})}\BibitemShut {NoStop}%
\bibitem [{\citenamefont {Giazotto}\ \emph {et~al.}(2006)\citenamefont
  {Giazotto}, \citenamefont {Heikkil\"a}, \citenamefont {Luukanen},
  \citenamefont {Savin},\ and\ \citenamefont {Pekola}}]{Giazotto06}%
  \BibitemOpen
  \bibfield  {author} {\bibinfo {author} {\bibfnamefont {F.}~\bibnamefont
  {Giazotto}}, \bibinfo {author} {\bibfnamefont {T.~T.}\ \bibnamefont
  {Heikkil\"a}}, \bibinfo {author} {\bibfnamefont {A.}~\bibnamefont
  {Luukanen}}, \bibinfo {author} {\bibfnamefont {A.~M.}\ \bibnamefont {Savin}},
  \ and\ \bibinfo {author} {\bibfnamefont {J.~P.}\ \bibnamefont {Pekola}},\
  }\href {\doibase 10.1103/RevModPhys.78.217} {\bibfield  {journal} {\bibinfo
  {journal} {Rev. Mod. Phys.}\ }\textbf {\bibinfo {volume} {78}},\ \bibinfo
  {pages} {217} (\bibinfo {year} {2006})}\BibitemShut {NoStop}%
\bibitem [{\citenamefont {Costi}\ and\ \citenamefont
  {Zlati\ifmmode~\acute{c}\else \'{c}\fi{}}(2010)}]{Costi10}%
  \BibitemOpen
  \bibfield  {author} {\bibinfo {author} {\bibfnamefont {T.~A.}\ \bibnamefont
  {Costi}}\ and\ \bibinfo {author} {\bibfnamefont {V.}~\bibnamefont
  {Zlati\ifmmode~\acute{c}\else \'{c}\fi{}}},\ }\href {\doibase
  10.1103/PhysRevB.81.235127} {\bibfield  {journal} {\bibinfo  {journal} {Phys.
  Rev. B}\ }\textbf {\bibinfo {volume} {81}},\ \bibinfo {pages} {235127}
  (\bibinfo {year} {2010})}\BibitemShut {NoStop}%
\bibitem [{\citenamefont {Jakobs}\ \emph {et~al.}(2007)\citenamefont {Jakobs},
  \citenamefont {Meden},\ and\ \citenamefont {Schoeller}}]{Jakobs07}%
  \BibitemOpen
  \bibfield  {author} {\bibinfo {author} {\bibfnamefont {S.~G.}\ \bibnamefont
  {Jakobs}}, \bibinfo {author} {\bibfnamefont {V.}~\bibnamefont {Meden}}, \
  and\ \bibinfo {author} {\bibfnamefont {H.}~\bibnamefont {Schoeller}},\ }\href
  {\doibase 10.1103/PhysRevLett.99.150603} {\bibfield  {journal} {\bibinfo
  {journal} {Phys. Rev. Lett.}\ }\textbf {\bibinfo {volume} {99}},\ \bibinfo
  {pages} {150603} (\bibinfo {year} {2007})}\BibitemShut {NoStop}%
\end{thebibliography}%
\end{document}